\pdfoutput=1
\documentclass[reprint,aps,pre,showpacs,twocolumn]{revtex4-1}
\usepackage{mathptmx} 
\usepackage{graphicx,color}
\usepackage{subfigure}
\usepackage{bm,amsmath,amssymb}
\usepackage{enumerate}

\newcommand{\eref}[1]{Eq.~(\ref{#1})}%
\newcommand{\Eref}[1]{Equation~(\ref{#1})}%
\newcommand{\fref}[1]{Fig.~\ref{#1}} %
\newcommand{\Fref}[1]{Figure~\ref{#1}}%
\newcommand{\sref}[1]{Sec.~\ref{#1}}%
\newcommand{\Sref}[1]{Section~\ref{#1}}%
\newcommand{\aref}[1]{Appendix~\ref{#1}}%

\begin{document}

\title{Work fluctuations for a Brownian particle in a 
  harmonic trap with fluctuating locations}

\author{Arnab Pal}
\author{Sanjib Sabhapandit}

\affiliation{Raman Research Institute, Bangalore 560080, India}

\date{\today}



\begin{abstract}
We consider a Brownian particle in a harmonic trap. The location of
the trap is modulated according to an Ornstein-Uhlenbeck process. We
investigate the fluctuation of the work done by the modulated trap on
the Brownian particle in a given time interval in the steady state. We
compute the large deviation as well as the complete asymptotic form of
the probability density function of the work done. The theoretical
asymptotic forms of the probability density function are in very good
agreement with the numerics. We also discuss the validity of the
fluctuation theorem for this system.
\end{abstract}
\maketitle

\section{Introduction}

Equilibrium statistical mechanics provides us a well-established
framework to deal with systems in thermal equilibrium.  When a system
is perturbed externally from its equilibrium state, the
so-called \emph{fluctuation-dissipation theorem} relates
the \emph{linear response} of the system (to the external
perturbation) to the \emph{fluctuations} properties of the system in
equilibrium (in the absence of the perturbation)~\cite{Kubo:1966}.
Within the framework of linear response theory, the Green-Kubo
relation gives linear transport coefficients in terms of integral over
time-correlation function of the corresponding current in
equilibrium~\cite{Green-Kubo}.  In contrast, a general understanding
of nonequilibrium (arbitrarily far from equilibrium) systems is rather
poor.  That is why there has been a lot of excitement surrounding
the \emph{fluctuation theorem}, which aims at making a general
statement about the fluctuations of entropy production during a
nonequilibrium process.  The fluctuation theorem has been suggested as
a natural extension of the fluctuation-dissipation theorem from the
linear response regime to arbitrarily far from equilibrium, as the
fluctuation theorem reduces to the Green-Kubo formula and the Onsager
reciprocity relations in the zero forcing limit~\cite{Gallavotti:96}.

Imagine, two heat reservoirs at different temperatures connected by a
thermal conductor. For a macroscopic object, we expect the heat to
flow across the conductor from the hotter to the colder reservoir, in
accordance with the second law of thermodynamics.  However, for small
systems, where microscopic fluctuations become important, once in a
while we might observe heat to flow from the colder to the hotter
reservoir. These reverse events are usually referred to as ``second
law violation''.  The fluctuation theorem gives a mathematical
expression for the ratio of the probability of ``obeying the second
law'' to that of the ``second law violation''.

Following a theoretical argument, Evans {\it et. al.} found a relation
between the probabilities of positive and negative entropy production
in the nonequilibrium steady state, in a molecular dynamics simulation
of a two-dimensional fluid driven by external shear and coupled to a
thermostat~\cite{Evans:93}.  Gallavotti and Cohen proved this relation
(and called it \emph{fluctuation theorem}) for the phase space
contraction (interpreted as the entropy production) for dissipative
dynamical systems in the nonequilibrium steady-state
using \emph{chaotic hypothesis} and time reversal
invariance~\cite{Gallavotti:95}.  Evans and Searles had derived
earlier a similar relation (now known  as the \emph{transient}
fluctuation theorem) for systems starting from equilibrium initial
condition~\cite{Evans:94}.  For stochastic systems, the fluctuation
theorem has been proven by Kurchan~\cite{Kurchan:98} for Langevin
dynamics and extended by Lebowitz and Spohn~\cite{Lebowitz:99} to
general Markov processes.
Subsequently, there has been an explosion of research activities
investigating the validity of the fluctuation theorem for other
quantities such as work, power flux, heat flow, total entropy, etc.,
both theoretically~\cite{Farago:02, vanZon:03, vanZon:04, Mazonka:99,
Narayan:04, Bodineau:04, Seifert:05, Baiesi:06, Bonetto:06, Visco:06,
Mai:07, Saito:07, Harris:07, Kundu:11, Saito:11, Sabhapandit:11-12,
Fogedby:11, Kundu:12} and experimentally~\cite{Wang:02, Wang:05,
Carberry:04, Goldburg:01, Feitosa:04, Garnier:05, Liphardt:02,
Collin:05, Majumdar:08, Douarche:06, Falcon:08, Bonaldi:09,
Ciliberto:10, Ciliberto:10-2, Gomez-Solano:11}.
The recent review~\cite{Seifert:2012} contains an extensive list of
references pointing to several other reviews as well as research
articles on fluctuation theorem and related topics.

Recently, Ref.~\cite{Ciliberto:10} reported experiments on the
fluctuations of the work done by an external Gaussian random force on
two different stochastic systems coupled to a thermal bath: (i) a
colloidal particle in an optical trap and (ii) an atomic-force
microscopy cantilever.  Analytical results have been obtained for the
second system in~\cite{Sabhapandit:11-12}.  In the first experiment, a
colloidal particle immersed in water (which acts as thermal bath) is
confined in an optical trap. The position of the trap is modulated
according to a Gaussian Ornstein-Uhlenbeck process. The authors have
experimentally determined the probability density function (PDF) of
the work done on the colloidal particle by the random force exerted by
the modulating trap. In this paper, we analytically treat this
problem.

The remainder of the paper is organized as follows.
In \sref{sec:model}, we define the model. \Sref{sec:GF} contains the
derivation of the moment generating function of work done $W_\tau$ in
a given time $\tau$, which has the form $\bigl \langle e^{-\lambda
W_\tau}\bigr\rangle
\approx g(\lambda)\,   e^{\tau\mu(\lambda)}$ for  large $\tau$.
In \sref{sec:PDF}, we invert the moment generating function to obtain
the asymptotic form (for large $\tau$) of the PDF of $W_\tau$.  We
find that in that relevant interval, $g(\lambda)$ can either be
analytic, or can have either one branch point or three or four branch
points, depending on the values of the tuning parameters of the
problem.  The case when $g(\lambda)$ is analytic, is simpler and the
asymptotic PDF can be obtained by the usual saddle point
approximation, which is given by \eref{saddle-point approximation}
in \sref{no-singularities}. In \sref{one-singularity}, we deal with
case when $g(\lambda)$ has one branch point. The cases when
$g(\lambda)$ has three and four branch points are discussed in
Secs. \ref{three-singularities} and \ref{four-singularities},
respectively.  The analytical results obtained in each section are
supported by numerical simulation performed on the
system. \Sref{sec:LDF} contains a discussion on large deviation
function and validity of the fluctuation theorem in the context of the
problem at hand. Finally, we summarize the paper
in \sref{sec:summary}.  Some details of calculation are pushed to two
appendices:
\aref{MGF calculation} contains the details of the calculation of the
moment generating function. In \aref{singularities of g}, we analyze
the singularities of $g(\lambda)$ and in
\aref{steepest-descent} we give the steepest descent method with
branch point to calculate the PDF.  We also provide an index of the
 relevant notations used in the paper in \aref{notations}, for the
 ease of quick look up.

\section{The model}
\label{sec:model}

Consider a Brownian particle suspended in a fluid at temperature $T$,
with the viscous drag coefficient $\gamma$. The particle is confined
in a quadratic potential (harmonic trap) around the position $y$ and
having a stiffness $k$.  The position $x(t)$ of the particle is
described by the overdamped Langevin equation
\begin{equation}
\frac{dx}{dt}=-\frac{ x -y}{\tau_\gamma}  +  \xi(t),
\label{Langevin-1}
\end{equation} 
where $\tau_\gamma=\gamma/k$ is the relaxation time of the harmonic
trap. The thermal noise $\xi(t)$ is taken to be Gaussian with mean
$\langle\xi(t)\rangle=0$ and covariance $\langle\xi(t)\xi(s)\rangle=2
D \delta(t-s)$, where the diffusion coefficient $D =
\gamma^{-1} k_B T$ with $k_B$ being the Boltzmann constant. An
external time-varying random force is exerted by the trap on the
Brownian particle by externally modulating the position of the trap
according to an Ornstein-Uhlenbeck process
\begin{equation}
\frac{dy}{dt}=-\frac{y}{\tau_{0}}  + \zeta(t),
\label{Langevin-2}
\end{equation}
where $\zeta(t)$ is an externally generated Gaussian white
(non-thermal) noise with mean $\langle \zeta(t)\rangle =0$ and
covariance $\langle \zeta(t) \zeta(s) \rangle = 2 A \delta(t-s)$.
There is no correlation between the externally applied noise and the
thermal noise, $\langle\zeta(t) \xi(s)\rangle=0$.  The system
eventually reaches steady state, and in the steady state the trap
exerts a correlated random force $k y(t)$ on the Brownian particle
with mean $\langle y(t)\rangle =0$ and covariance $\langle y(t)
y(s)\rangle = A \tau_0 \exp(-|t-s|/\tau_0)$. The quantity of our
interest is the work done in the steady state, by the random force
exerted by the trap on the Brownian particle in a given time duration
$\tau$.  This is given (in units of $k_{B}T$) by
\begin{equation}
W_{\tau}=\frac{1}{k_{B}T}\int_{0}^{\tau}  k y(t)\, \frac{dx}{dt}\,
dt,
\label{work}
\end{equation}
with the initial condition (at $\tau=0$) drawn from the steady state
distribution.

It is convenient to use the following dimensionless parameters
\begin{equation}
\theta = A/D,\quad\text{and}\quad
\delta = \tau_0/\tau_\gamma.
\label{theta-delta}
\end{equation}
From an experimental perspective~\cite{Ciliberto:10}, it is natural to
use another parameter that measures the deviation of the system from
equilibrium:
\begin{equation}
\alpha=\frac{\langle x^{2}\rangle}{\langle x^{2}\rangle_\text{eq}} - 1,
\label{alpha}
\end{equation}
where $\langle x^2\rangle$ is the variance of $x$ in the steady state
in the presence of trap modulation, whereas $\langle
x^2\rangle_\text{eq}=D\tau_\gamma$ is the corresponding variance at
equilibrium, i.e., without the presence of the trap modulation
($y=0$). It should be noted that, the three parameters introduced
above are not independent of each others and are related by
\begin{equation}
\alpha=\theta\delta^2(1+\delta)^{-1}.
\label{alpha-theta-delta}
\end{equation}

The mean work can be computed easily using the above equations and one
finds $\langle W_\tau\rangle\approx \alpha\tau/\tau_0$ for large
$\tau$. Although the mean work is positive (and large for large
$\tau$), there can be negative fluctuations (with small probabilities)
and the fluctuation theorem quantifies the ratio of the probabilities
of the positive and the negative fluctuations. However, the aim of
this paper is not to merely check whether the fluctuation theorem is
satisfied or not for this system, but to obtain the distribution of
the work done (for large $\tau$), which contains much more information
about the system than the former.

\section{Moment Generating Function}
\label{sec:GF}

To compute the distribution of $W_\tau$, we first consider the moment
generating function restricting to fixed initial and final
configurations $(x_0,y_0)$ and $(x,y)$ respectively:
\begin{equation}
Z(\lambda,x,y,\tau|x_0,y_0)=\bigl \langle e^{-\lambda
W_\tau}\, \delta[x-x(\tau)] \delta[y-y(\tau)] \bigr\rangle_{(x_0,y_0)},
\label{restricted GF}
\end{equation}
where $\langle \cdots\rangle_{(x_0,y_0)}$ denotes an average over the
histories of the thermal noises starting from the initial condition
$(x_0,y_0)$.  It can be shown that $Z(\lambda,x,y,\tau|x_0,y_0)$
satisfies the Fokker-Planck equation
\begin{equation}
\frac{\partial Z}{\partial \tau} 
= \mathcal{L}_\lambda Z
\label{FP-eq}
\end{equation}
with the initial condition
$Z(\lambda,x,y,0|x_0,y_0)= \delta(x-x_0)\, \delta(y-y_0)$, and the
Fokker-Planck operator is given by
\begin{multline}
\mathcal{L_{\lambda}}= 
D\frac{\partial^2}{\partial x^2}
+\theta D\frac{\partial^2}{\partial y^2} 
+\frac{\delta}{\tau_0}\frac{\partial}{\partial x} (x-y) 
+\frac{1}{\tau_0}\frac{\partial}{\partial y} y \\
+\frac{2\lambda\delta}{\tau_0} y\frac{\partial}{\partial x}
+\frac{\lambda\delta^2}{\tau_0^2D}y (x-y)
+\frac{\lambda^2\delta^2}{\tau_0^2D}y^2.
\label{langevin operator}
\end{multline}
We do not know whether the above partial differential equation can be
solved to obtain $Z$. Fortunately, however, one does not require the
complete solution of the above equation to determine the large-$\tau$
behavior of the distribution of $W_\tau$.

The solution of the Fokker-Planck equation can be formally expressed
in the eigenbases of the operator $\mathcal{L}_\lambda$ and the large
$\tau$ behavior is dominated by the term having the largest
eigenvalue. Thus, for large-$\tau$,
\begin{equation}
  Z(\lambda,x,y,\tau |x_0,y_0)
= \chi(x_0,y_0,\lambda)\Psi(x,y,\lambda)\, e^{\tau\mu(\lambda)}
+\dotsb, \label{characteristic.1}
\end{equation}
where $\Psi(x,y,\lambda)$ is the eigenfunction corresponding to the
largest eigenvalue $\mu(\lambda)$ and $\chi(x_0,y_0,\lambda)$ is the
projection of the initial state onto the eigenstate corresponding to
the eigenvalue $\mu(\lambda)$. While we cannot solve the Fokker-Planck
equation, the functions in \eref{characteristic.1} can be obtained
using a method developed~\cite{Kundu:11} and
used~\cite{Sabhapandit:11-12} recently. We sketch the derivation in
the context of the present problem in \aref{MGF calculation}, where we
find
\begin{equation}
\mu(\lambda)=\frac{1}{2 \tau_{c}}[1-\nu(\lambda)],\quad
\tau_{c} = \tau_0(1+\delta)^{-1},
\label{mu}
\end{equation}
in which $\nu(\lambda)$ is given by,
\begin{equation}
\nu(\lambda)= \sqrt{1+4a \lambda(1-\lambda)}, \quad
a=\alpha(1+\delta)^{-1}.
\label{nu-1}
\end{equation}
We observe that the eigenvalue satisfies the Gallavotti-Cohen symmetry
$\mu(\lambda)=\mu(1-\lambda)$. In terms of the column vector
$U=(x,y)^T$, the eigenfunctions are
\begin{align}
&\Psi(x,y,\lambda)=\frac{1}{2\pi \sqrt{\det {H}_{1}(\lambda)}} \exp\left[-\frac{1}{2}U^{T}{L}_{1}(\lambda)U\right], \\
&\chi(x_0,y_0,\lambda)=\exp\left[-\frac{1}{2}U_{0}^{T}{L}_{2}(\lambda)U_{0}\right],
\end{align}
where the matrices ${H}_1$, ${L}_{1}$, and ${L}_{2}$ are given
in \aref{MGF calculation}.

Using the explicit forms one can verify the eigenvalue equation
$\mathcal{L}_\lambda \Psi (x,y,\lambda)
= \mu(\lambda) \Psi(x,y,\lambda)$. Moreover,
\begin{math}
\int_{-\infty}^\infty\int_{-\infty}^\infty
\chi(x,y,\lambda) \Psi(x,y,\lambda)\, dx\, dy =1,
\end{math}
as expected. From the above expressions, we also find that $\mu(0)=0$
and $\chi(x_0,y_0,0)=1$.  Since the $\lambda=0$ case
of \eref{restricted GF} gives the PDF of the variables $(x,y)$ and
$\mu(0)$ is the largest eigenvalue, it follows
from \eref{characteristic.1} that $\Psi(x,y,0)$ is the steady-state
PDF of $(x,y)$.  Therefore, averaging over the initial variables
$(x_0,y_0)$ with respect to the steady-state PDF $\Psi(x_0,y_0,0)$ and
integrating over the final variables $(x,y)$, we find the moment
generating function of the work in the steady state as
\begin{equation}
Z(\lambda,\tau) =\bigl \langle e^{-\lambda W_\tau}\bigr\rangle
= g(\lambda)\,   e^{\tau\mu(\lambda)}+\dotsb,
\label{Z-asymptotic}
\end{equation}
where
\begin{align}
g(\lambda)=\frac{2}{\sqrt{\nu(\lambda)+1-2b_{+}\lambda}\,
\sqrt{\nu(\lambda)+1-2b_{-}\lambda}}\notag\\ 
\times 
\frac{2\nu(\lambda)}{\sqrt{\nu(\lambda)+1+2b_{+}\lambda}\,
\sqrt{\nu(\lambda)+1+2b_{-}\lambda}},
\label{g}
\end{align}
with
\begin{equation}
b_{\pm}=\frac{\alpha}{2}\left[1\pm\sqrt{1+\frac{4}{\theta\delta}}\right].
\label{b_pm}
\end{equation}
The first factor in the above expression of $g(\lambda)$ is due to the
averaging over the initial conditions with respect to the steady-state
distribution and the second factor is due to the integrating out of
the final degrees of freedom.

\section{Probability Density Function}
\label{sec:PDF}

The PDF of the work done $W_\tau$ can be obtained from the moment
generating function $Z(\lambda,\tau)$, by taking the inverse
``Fourier'' (two-sided Laplace) transform
\begin{equation}
P(W_\tau) = \frac{1}{2\pi i} \int_{-i\infty}^{+i\infty} Z(\lambda,\tau)\,
e^{\lambda W_\tau}\, d\lambda,
\end{equation}
where the integration is done along the imaginary axis in the complex
$\lambda$ plane. Using the large-$\tau$ form of $Z(\lambda,\tau)$
given by \eref{Z-asymptotic} we write
\begin{equation}
P(W_{\tau}=w\tau)\approx \frac{1}{2\pi i} \int_{-i\infty}^{+i\infty}
g(\lambda) e^{\tau f_w(\lambda)} d\lambda,
\label{P(W)}
\end{equation}
where 
\begin{equation}
f_w(\lambda)= \frac{1}{2}\bigl[1-\nu(\lambda) \bigr]
+\lambda w.
\label{f_w}
\end{equation}
and we have set $\tau_c=1$ for convenience. This is completely
equivalent to measuring the time in the unit of $\tau_c$, that is,
$\tau/\tau_c \rightarrow
\tau$.

The large-$\tau$ form of $P(W_\tau)$ can be obtained from \eref{P(W)}
by using the method of steepest descent.  The saddle-point $\lambda^*$
is obtained from the solution of the condition $f'_w(\lambda^*)=0$ as
\begin{equation}
\lambda^*(w) =\frac{1}{2}
\left[1-\frac{w}{\sqrt{w^2+a}}\sqrt{1+\frac{1}{a}} \right].
\label{lambda*}
\end{equation}
From the above expression one finds that $\lambda^*(w)$ is a
monotonically decreasing function of $w$ and
$\lambda^*(w\rightarrow\mp \infty)\rightarrow\lambda_\pm$, where
\begin{equation}
\lambda_\pm =\frac{1}{2}\left[1\pm \sqrt{1+\frac{1}{a}} \right].
\label{lambda_pm}
\end{equation}
Therefore, $\lambda^*\in (\lambda_-,\lambda_+)$.  It is also useful to
note that $\nu(\lambda)$ can be written in terms of $\lambda_\pm$ as
\begin{equation}
\nu(\lambda)=\sqrt{4a(\lambda_+-\lambda)(\lambda-\lambda_-)}.
\label{nu-2}
\end{equation}
This clearly shows that $\nu(\lambda)$ has two branch points on the
real-$\lambda$ line at $\lambda_{\pm}$. However, $\nu(\lambda)$ is
real and positive in the (real) interval
$\lambda\in(\lambda_{-}, \lambda_{+})$. As a consequence,
$f_w(\lambda)$ remains real in the interval
$(\lambda_{-}, \lambda_{+})$. At $\lambda=\lambda^*$ we find
\begin{equation}
\nu(\lambda^*)=\frac{\sqrt{a(1+a)}}{\sqrt{w^2+a}},
\label{nu-lambda-star}
\end{equation}
and 
\begin{equation}
h_\text{s}(w):=f_w(\lambda^*) =\frac{1}{2}\left[1+w
- \sqrt{w^2+a}\sqrt{1+\frac{1}{a} }\right].
\label{hs(w)}
\end{equation}
One also finds that
\begin{equation}
f''_w (\lambda^*)= \frac{2
(w^2+a)^{3/2}}{\sqrt{a(1+a)}} >0.
\label{f2}
\end{equation}
This means that $f_w(\lambda)$ has a minimum at $\lambda^*$ along
real-$\lambda$, and hence the path of steepest descent is perpendicular
to the real-$\lambda$ axis at $\lambda=\lambda^*$.

\begin{figure*}
\includegraphics[width=.45\hsize]{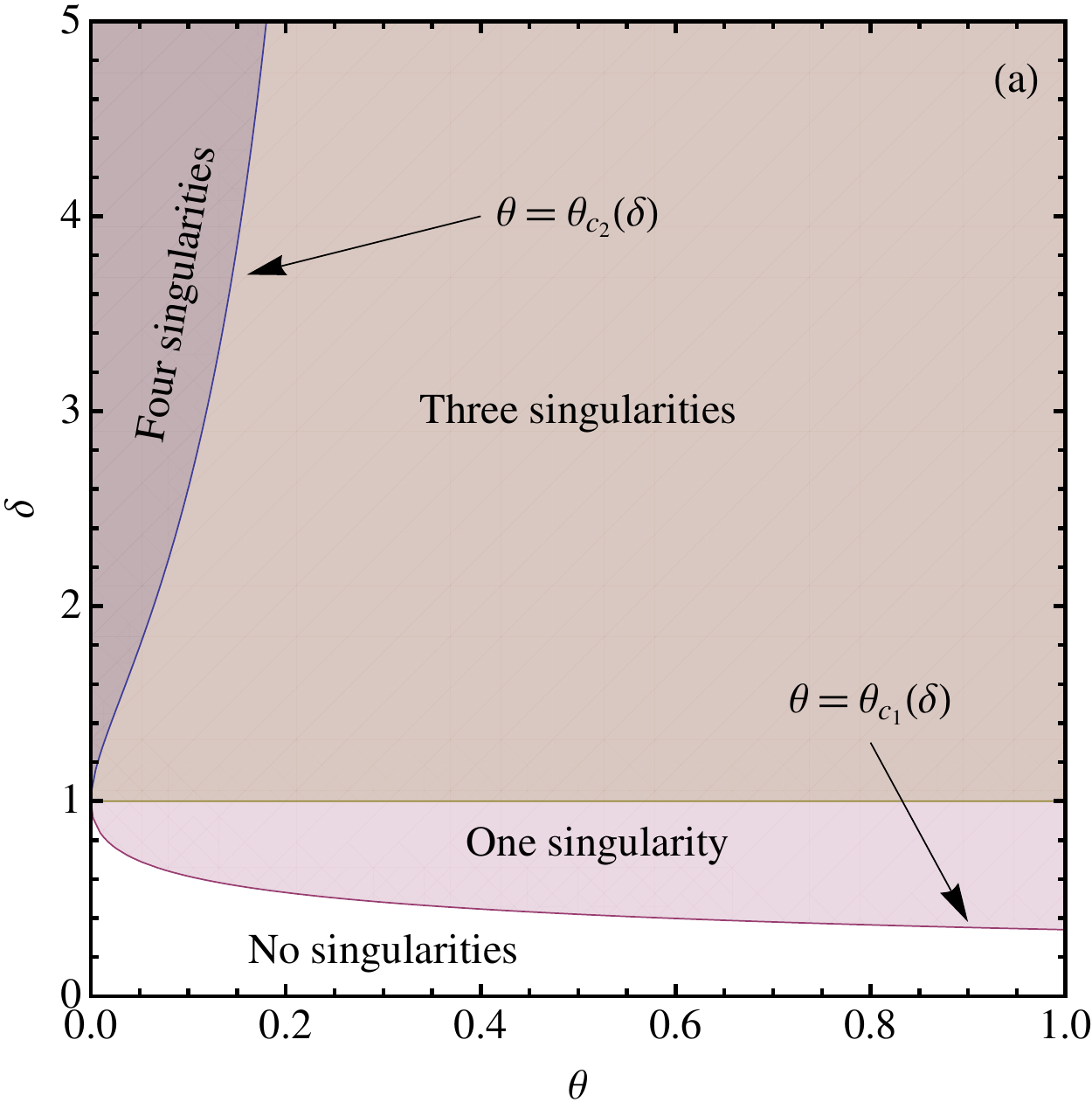}~
\includegraphics[width=.45\hsize]{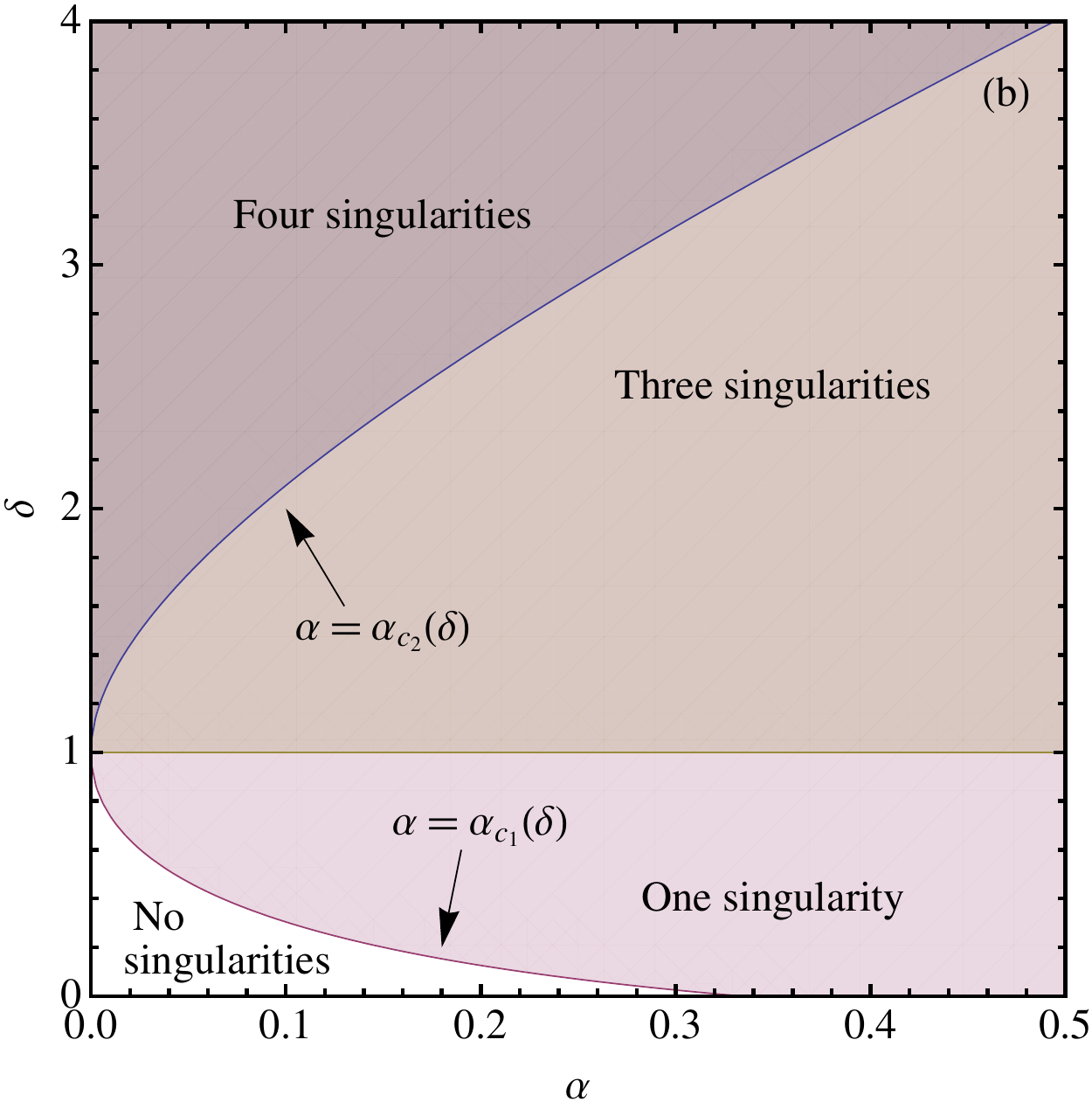}
\caption{\label{phase2}(Color online)  The regions in the
(a) $\theta,\delta$ and (b) $\alpha,\delta$ spaces, where $g(\lambda)$
has the number of singularities mentioned in the figure. The equations
of the boundary lines separating different regions are given
in \aref{singularities of g}. $\alpha_{c_1}=1/3$ for $\delta=0$ and
$\theta_{c_2}\rightarrow 1/3$ as $\delta\rightarrow\infty$.  Each of
the phase boundaries meet at $\theta=0$ ($\alpha=0$), $\delta=1$.}
\end{figure*}

Now, if $g(\lambda)$ is analytic for $\lambda\in(0,\lambda^*)$, one
can deform the contour along the path of the steepest descent through
the saddle-point, and obtain $P(W_\tau)$ using the usual saddle-point
approximation method.  However, if $g(\lambda)$ has any singularities,
then the straightforward saddle-point method cannot be used, and one
would require more sophisticated methods to obtain the asymptotic form
of $P(W_\tau)$. Therefore, it is essential to analyze $g(\lambda)$ for
possible singularities.  In \aref{singularities of g}, we examine the
terms under the four square roots in the denominator of $g(\lambda)$
in \eref{g}.

In \fref{phase2}, we show the regions in the $(\theta,\delta)$ and
$(\alpha,\delta)$ planes, where $g(\lambda)$ possesses singularities.

\subsection{The case of no singularities}
\label{no-singularities}

\begin{figure}
\includegraphics[width=\hsize]{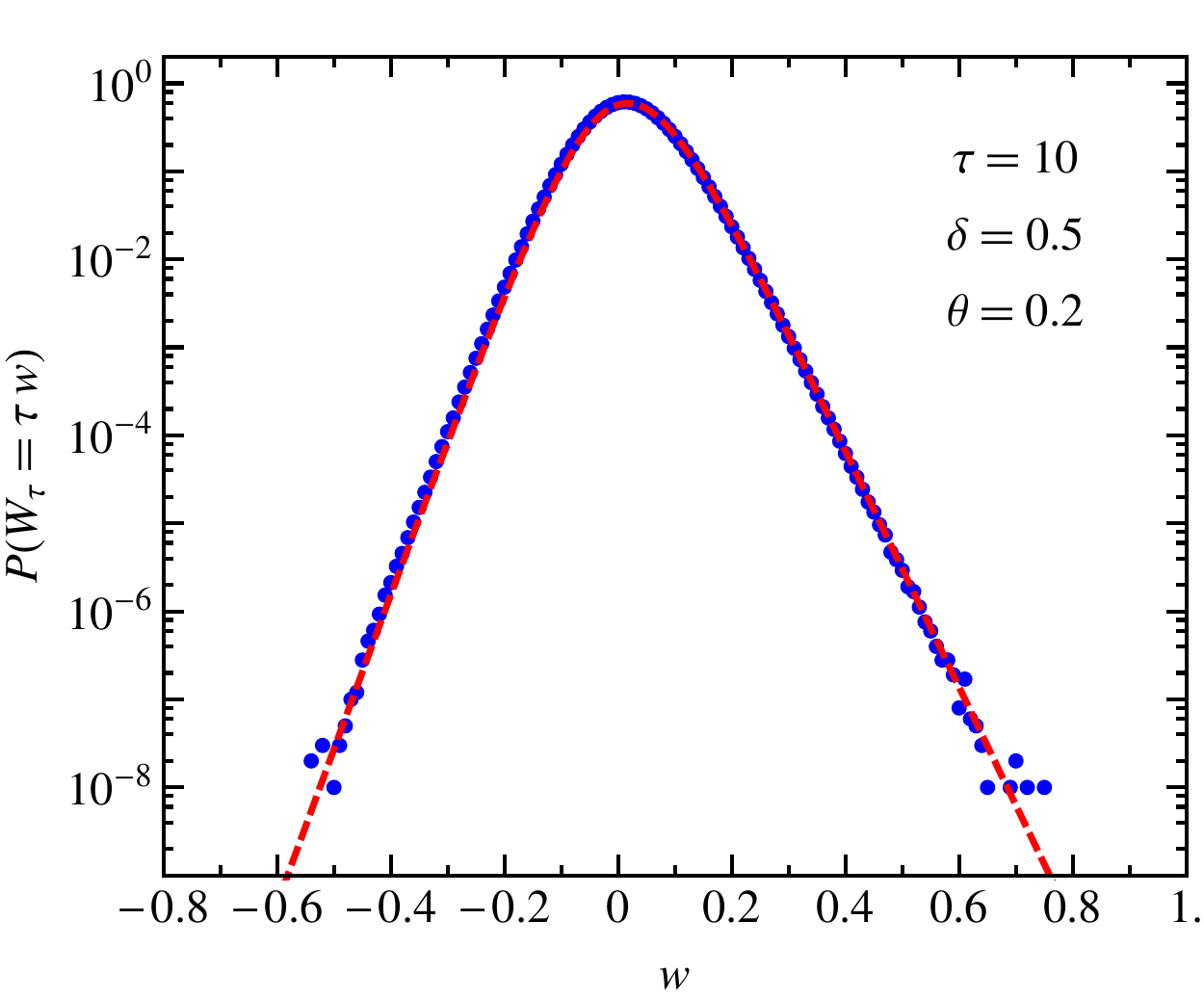}
\caption{\label{pdf0} (Color online). $P(W_\tau)$ against the scaled
variable $w=W_\tau/\tau$ for $\tau=10$, $\tau_c=1$. The points (blue)
 are obtained from numerical simulation, and the dashed solid lines
 (red) plot the analytical asymptotic forms given
 by \eref{saddle-point approximation}.
 $\theta_{c_1}=9/35=0.257\dotsc$ for $\delta=1/2$.}
\end{figure}

In the singularity free region $\delta<1$, $\theta<\theta_{c_1}$
($\alpha <\alpha_{c_1}$), the asymptotic PDF of the work done is
obtained by following the usual saddle-point approximation method.  We
get
\begin{equation}
P(W_{\tau}=w\tau)\approx
\frac{g(\lambda^*) e^{\tau h_\text{s}(w)}}
{\sqrt{2\pi\tau f''_w(\lambda^*)}},
\label{saddle-point approximation}
\end{equation}
where $ h_\text{s}(w)$ and $f''_w(\lambda^*)$ are given by
Eqs.~\eqref{hs(w)} and \eqref{f2}, respectively, and $g(\lambda^*)$
can be obtained from \eref{g} while using $\lambda^*$
from \eref{lambda*}.  \Fref{pdf0} shows very good agreement between
the form given by \eref{saddle-point approximation} and numerical
simulation results for $\theta < \theta_{c_1}$.

\subsection{The case of one singularity}
\label{one-singularity}

In the case $\delta <1$, $\theta>\theta_{c_1}$ where $g(\lambda)$ has
only one singularity, or the case $\delta=1$ where only one
singularity of $g(\lambda)$ is relevant, we can write
\begin{equation}
g(\lambda)=
\frac{g_1(\lambda)}{\sqrt{\lambda_\text{a}-\lambda}},
\label{g-lambda-1}
\end{equation}
where $g_1(\lambda)$ is the analytical factor of $g(\lambda)$.

It is evident that for a given value of $\delta$ and $\theta$, the
position of the branch point $\lambda_\text{a}$ is fixed somewhere
between the origin and $\lambda_+$. On the other hand, according
to \eref{lambda*}, even for a fixed $\theta$, the saddle-point
$\lambda^*(w)$ moves unidirectionally along the real-$\lambda$ line
from $\lambda_-$ to $\lambda_+$ as one decreases $w$ from $+\infty$ to
$-\infty$ in a monotonic manner. Therefore, for sufficiently large
$w$, the saddle-point lies in the interval
$(\lambda_-,\lambda_\text{a})$, and therefore, the contour of
integration in \eref{P(W)} can be deformed into the steepest descent
path (that passes through $\lambda^*$) without touching
$\lambda_\text{a}$ (see \fref{steepest-descent contour1a}). However,
as one decreases $w$, the saddle-point hits the branch-point,
$\lambda^*(w^*_\text{a})=\lambda_\text{a}$, at some specific value
$w=w^*_\text{a}$ given by \eref{wstar}. For $w < w^*_\text{a}$, since
$\lambda^* > \lambda_\text{a}$, the steepest descent contour wraps
around the branch-cut between $\lambda_\text{a}$ and $\lambda^*$ as
shown in~\fref{steepest-descent contour1b}. Leaving the details of the
calculation to \aref{steepest-descent}, here we present the main
results.

\subsubsection{$w> w^*_\mathrm{a}$}

For $w> w^*_\text{a}$, following \aref{sec: branch point outside},  we
get
\begin{equation}
P(W_{\tau}=w\tau)\approx
\frac{g(\lambda^*) e^{\tau h_\text{s}(w)}}
{\sqrt{2\pi\tau f''_w(\lambda^*)}}\,
R_1\biggl(\sqrt{\tau \bigl[h_\text{a}(w)-h_\text{s}(w)\bigr]}\biggr),
\label{P(W)-1a}
\end{equation}
where the function $R_1(z)$ is given by
\begin{equation}
R_1(z)=\frac{z}{\sqrt{\pi}}\,  e^{z^2/2}\, K_{1/4} (z^2/2),
\label{R_1}
\end{equation}
with $ K_{1/4} (z)$ being the modified Bessel function of the second
kind.  It follows from the asymptotic form of $K_{1/4}(z)$ that
$R_1(z\rightarrow\infty)\rightarrow 1$. Therefore, for $w\gg
w^*_\text{a}$, \eref{P(W)-1a} approaches the form of the usual
saddle-point approximation given by \eref{saddle-point
approximation}. On the other hand, using $K_{1/4}(z)\simeq
(1/2)\Gamma(1/4)(z/2)^{-1/4}$ for small $z$, we get
$R_1(z)\simeq \Gamma(1/4) \sqrt{z/2\pi}$.  As $w \rightarrow
w^*_\text{a}$ from above, i.e., when the saddle point approaches the
branch point from below, $h_\text{a}(w)-h_\text{s}(w)\equiv
f_w(\lambda_\text{a})-f_w(\lambda^*) \simeq
(\lambda_\text{a}-\lambda^*)^2 f''(\lambda^*)/2$. Therefore, the
expression given by \eref{P(W)-1a} remains finite, even when the
saddle point approaches the singularity, i.e.,
\begin{equation}
P(W_{\tau}=w\tau)\approx
\frac{\Gamma(1/4)}{2\pi}\frac{g_1(\lambda^*) e^{\tau h_\text{s}(w)}}{[2\tau f''_w(\lambda^*)]^{1/4}}\,\quad\text{as} ~w\rightarrow
w^*_\text{a}.
\label{P(W)-1a*}
\end{equation}

\subsubsection{$w < w^*_\mathrm{a}$}

For $w< w^*_\text{a}$, following \aref{sec: branch point inside}, we
write
\begin{equation}
P(W_{\tau}=w\tau)\approx P_\text{B} (w,\tau) + P_\text{S} (w,\tau),
\label{P(W)-1b}
\end{equation}
where $P_\text{B} (w,\tau)$ is the contribution coming from the
integrations along the branch cut and $P_\text{S}(w,\tau)$ is the
saddle point contribution. Following \aref{sec: Branch cut
contribution} we get,
\begin{equation}
P_\text{B}(w,\tau) \approx
\frac{\widetilde{g}(\lambda_\text{a})\, e^{\tau
h_\text{a}(w)}}{\sqrt{\pi\tau|f'_w(\lambda_\text{a})|}} 
R_3 
\left(\sqrt{\tau \bigl[h_\text{a}(w)-h_\text{s}(w)\bigr]}\right),
\label{P(W)-1b1}
\end{equation}
where
\begin{equation}
R_3(z)=\sqrt\frac{2z}{\pi}\, R_2(z) ,
\label{R2tilde}
\end{equation}
with $R_2(z)$ being given by \eref{R2}. Using the asymptotic forms of
$R_2(z)$ given in \aref{sec: Branch cut contribution}, we get
$R_3(z)\rightarrow 1$ in the limit $z\rightarrow\infty$. Therefore,
\begin{equation}
P_\text{B}(w,\tau) \sim \frac{\widetilde{g}(\lambda_\text{a})e^{\tau
h_\text{a}(w)}}{\sqrt{\pi\tau|f'_w(\lambda_\text{a})|}} \,
\quad\text{for}~~ w \ll w^*_\text{a}.
\end{equation}
As $w\rightarrow w^*_\text{a}$ (from below),
$P_\text{B}(w,\tau)\rightarrow 0$.

The contribution coming from the saddle point is given by
(see \aref{sec: Saddle point contribution}),
\begin{equation}
P_\text{S} (w,\tau) \approx
\frac{|g(\lambda^*)| e^{\tau h_\text{s}(w)}}
{\sqrt{2\pi\tau f''_w(\lambda^*)}}\,
R_4\biggl(\sqrt{\tau \bigl[h_\text{a}(w)-h_\text{s}(w)\bigr]}\biggr),
\label{P(W)-1b2}
\end{equation}
where the function $R_4(z)$ is given by
\begin{align}
R_4(z)=&\sqrt{\frac{\pi}{2}}\,z\, e^{z^2/2}\bigl[I_{-1/4} (z^2/2) + I_{1/4}
(z^2/2) \bigr]\notag\\ 
&- \frac{4z}{\sqrt\pi} \,{}_2F_2\left(1/2,1;3/4,5/4;z^2\right),
\label{R4}
\end{align}
where $I_{\pm1/4} (z)$ are modified Bessel functions of the first kind
and ${}_2F_2(a_1,a_2;b_1,b_2;z)$ is the generalized hypergeometric
function, defined by \eref{2F2}. The small and large $z$ behaviors of
$R_4(z)$ are given in \aref{sec: Saddle point contribution}.

For $w\ll w^*_\text{a}$ we get $P_\text{S} (w,\tau) \ll P_\text{B}
(w,\tau)$. On the other hand $P_\text{S}(w,\tau)$ acquires the same
limiting form as in \eref{P(W)-1a*}, when $w\rightarrow w^*_\text{a}$
(from below).

\subsubsection{Numerical Simulation}

\begin{figure}
\includegraphics[width=\hsize]{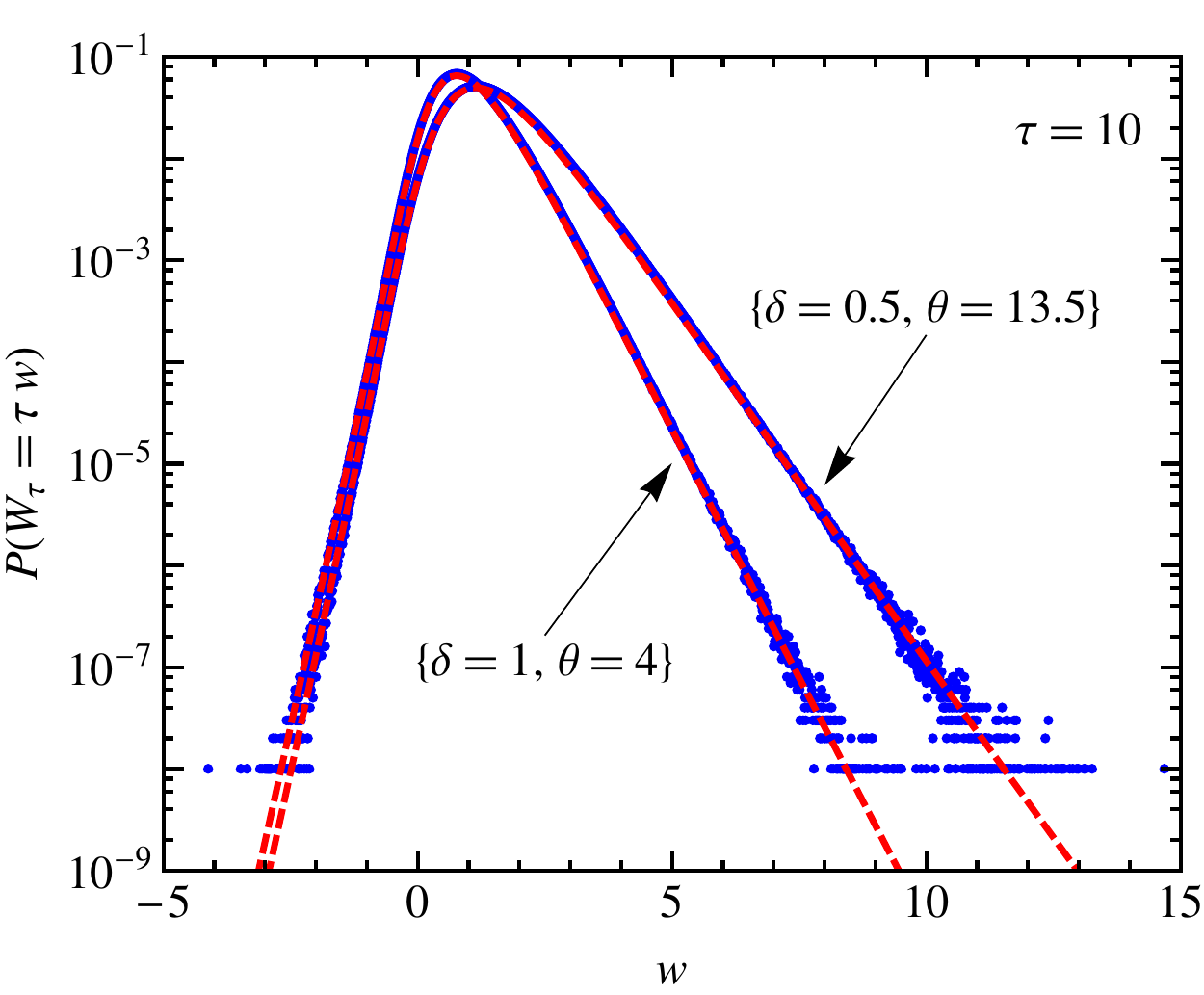}
\caption{\label{pdf1} (Color online). $P(W_\tau)$ against the scaled
variable $w=W_\tau/\tau$ for $\tau=10$, $\tau_c=1$. The points (blue)
 are obtained from numerical simulation, and the dashed solid lines
 (red) plot the analytical asymptotic forms given by \eref{P(W)-1a}
 for $w > w^*_\text{a}$ and Eqs.~\eqref{P(W)-1b}--\eqref{P(W)-1b2} for
 $w< w^*_\text{a}$, where $w^*_\text{a}=0$ for $\delta=1, \theta=4$, 
 and  $w^*_\text{a}=-0.0135\dotsc$ for $\delta=0.5, \theta=13.5$.}
\end{figure}

We now compare the asymptotic forms presented in this subsection with
numerical simulation.  In one case, we choose $\delta=1$ and
$\theta=4$, for which we get $\lambda_\pm=(1\pm\sqrt{2})/2$,
$\lambda^*(w)=\bigl(1-\sqrt2 w/\sqrt{1+w^2}\bigr)/2$,
$\lambda_\text{a}=1/2$, and $w^*_\text{a}=0$. In an another case, we
choose $\delta=1/2$ and $\theta=13.5$, for which
$w^*_\text{a}=-0.0135\dotsc$.
\Fref{pdf1} shows very good agreement between the
 analytical and and simulation results.

\begin{figure}
\includegraphics[width=.48\hsize]{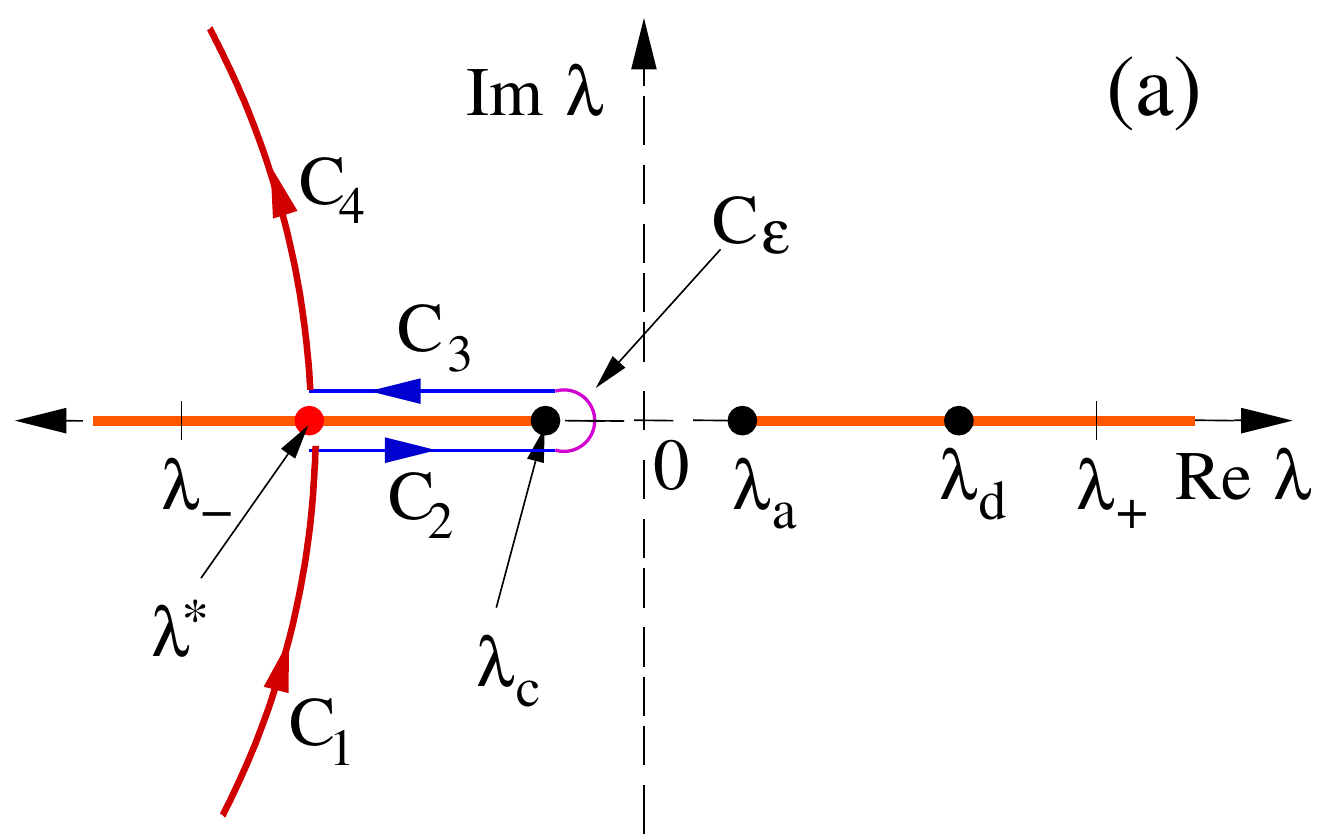}\hfill
\includegraphics[width=.48\hsize]{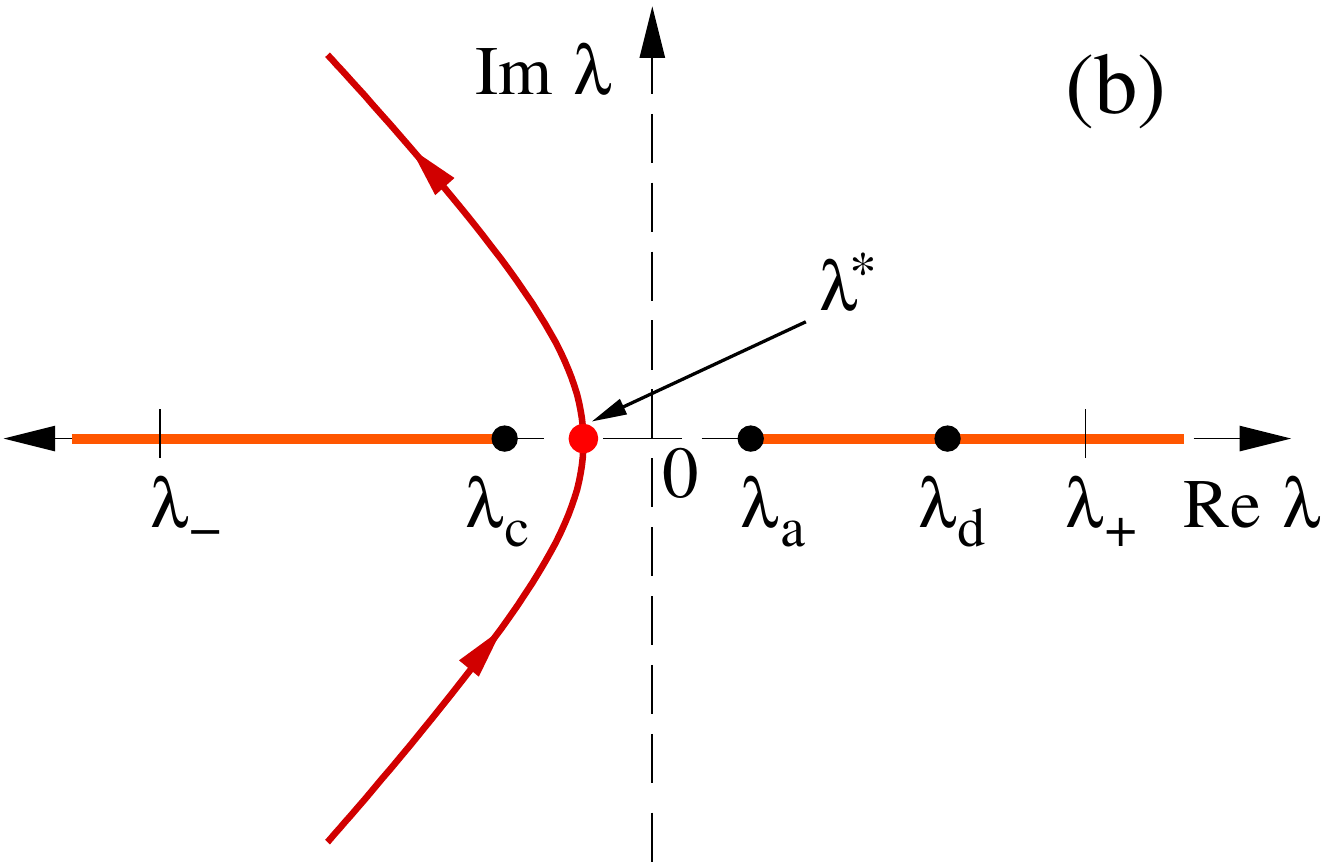}\\[2mm]
\includegraphics[width=.48\hsize]{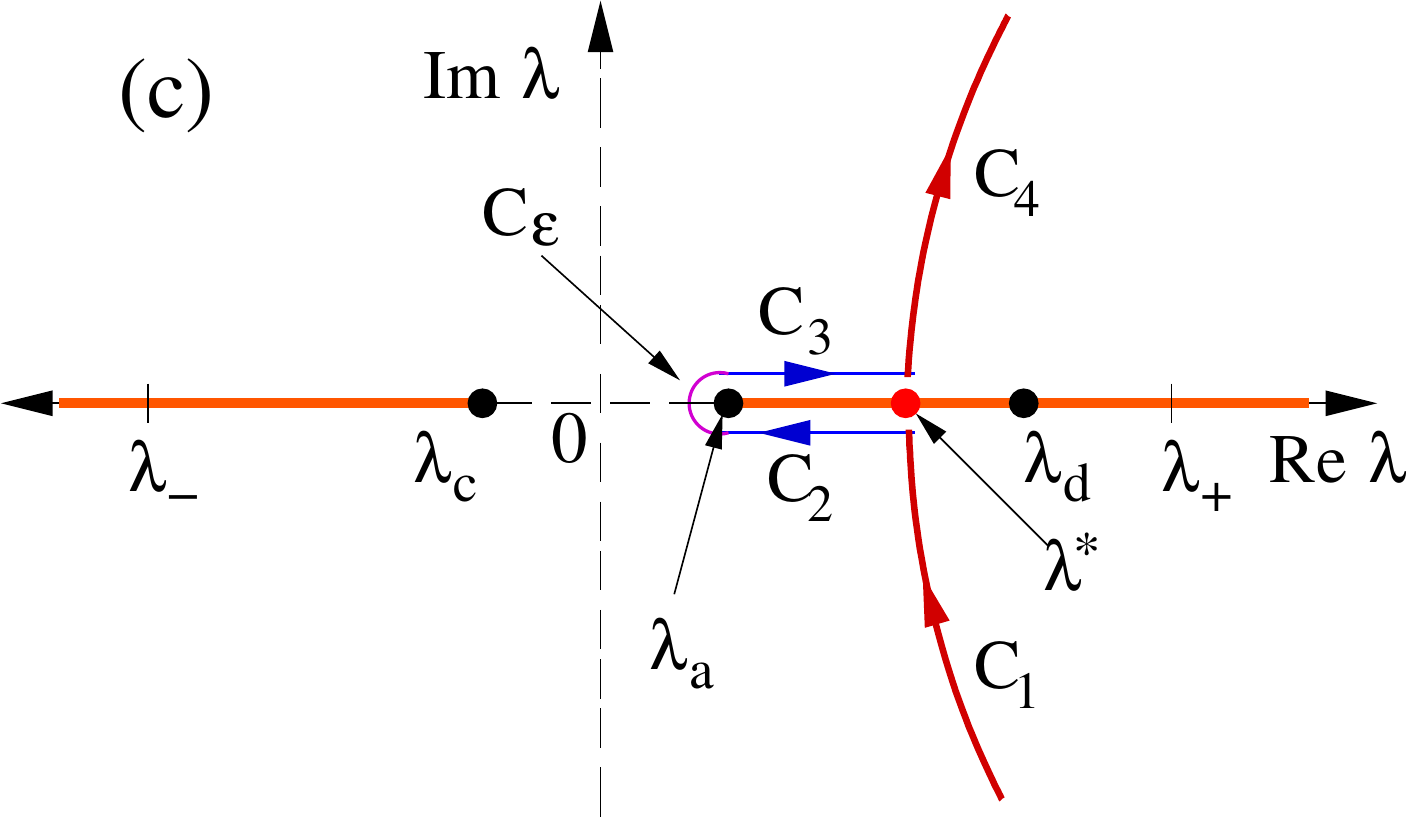}\hfill
\includegraphics[width=.48\hsize]{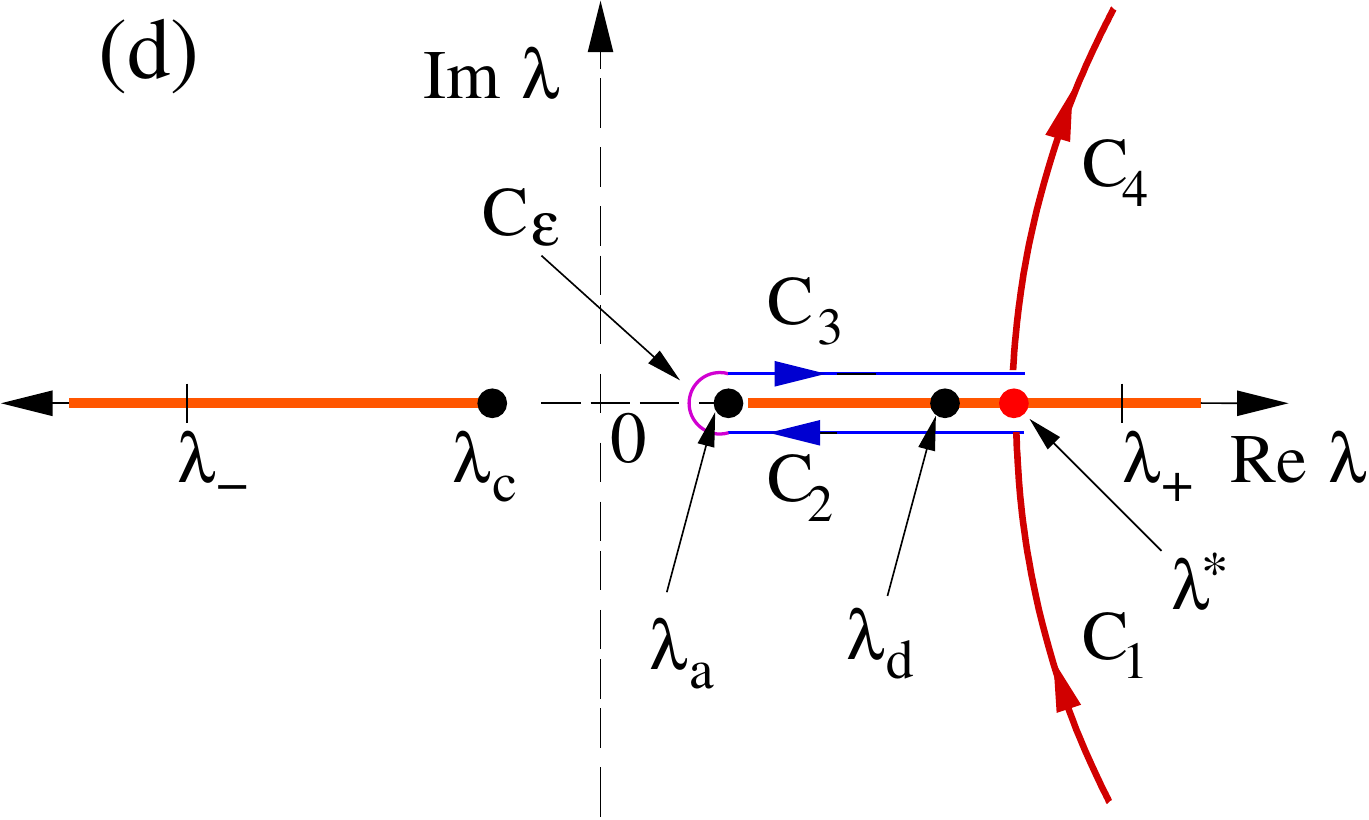}
\caption{\label{steepest-descent contour3} (Color online) 
Schematic steepest descent contours for the case when there are three
branch points at $\lambda_\text{a}$, $\lambda_\text{c}$ and
$\lambda_\text{d}$, where $\lambda_- < \lambda_\text{c} < 0
< \lambda_\text{a} < \lambda_\text{d}<\lambda_+$; and the saddle point
$\lambda^*$ lies between (a) $\lambda_-$ and $\lambda_\text{c}$, (b)
$\lambda_\text{c}$ and $\lambda_\text{a}$, (c) $\lambda_\text{a}$ and
$\lambda_\text{d}$, and (d) $\lambda_\text{d}$ and $\lambda_+$
respectively. }
\end{figure}

\subsection{The case of three singularities}
\label{three-singularities}

Now we consider the case, $\delta >1$ and $\theta > \theta_{c_2}$, in
which case $g(\lambda)$ has three singularities (see \fref{phase2}) at
$\lambda_\text{a}$, $\lambda_\text{c}$ and $\lambda_\text{d}$ given by
Eqs. \eqref{lambda-a}, \eqref{lambda-c} and \eqref{lambda-d}
respectively; where $\lambda_- < \lambda_\text{c} < 0
< \lambda_\text{a} < \lambda_\text{d}<\lambda_+$.  Therefore,
$g(\lambda)$ can be written as
\begin{equation}
g(\lambda)=\frac{g_3(\lambda)}{
\sqrt{\lambda-\lambda_\text{c}}
\sqrt{\lambda_\text{a}-\lambda}
\sqrt{\lambda_\text{d}-\lambda}},
\end{equation}
where $g_3(\lambda)$ is the analytical factor of $g(\lambda)$.  We
notice from \eref{lambda*} that $\lambda^*\rightarrow\lambda_-$ as
$w\rightarrow+\infty$ and $\lambda^*$ increases monotonically towards
$\lambda_+$ with decreasing $w$. Therefore, there are specific values
$+\infty > w^*_\text{c} > w^*_\text{a} > w^*_\text{d} >-\infty$ of $w$
given by \eref{wstar} at which the saddle point hits the corresponding
branch point, i.e., $\lambda^*(w^*_\text{c})=\lambda_\text{c}$,
$\lambda^*(w^*_\text{a})=\lambda_\text{a}$ and
$\lambda^*(w^*_\text{d})=\lambda_\text{d}$.

\begin{widetext}

\subsubsection{$w > w^*_\text{c}$}

For $w > w^*_\text{c}$, the saddle point lies between $\lambda_-$ and
$\lambda_\text{c}$. Therefore, as in the case of one singularity
discussed above in \sref{one-singularity}, the contributions comes
from the branch point as well as from the saddle point, as shown
in \fref{steepest-descent contour3}~(a). Following the procedure
similar to that in the one singularity case (see \aref{sec: branch
point inside}), we get
\begin{align}
P(W_\tau=w\tau)\approx &
\frac{\widetilde{g}(\lambda_\text{c})\,e^{\tau h_\text{c}(w)}}
{\sqrt{\pi\tau|f'_w(\lambda_\text{c})|}} 
R_3 
\left(\sqrt{\tau \bigl[h_\text{c}(w)-h_\text{s}(w)\bigr]}\right)
\notag \\
+&\frac{|g(\lambda^*)| e^{\tau h_\text{s}(w)}}
{\sqrt{2\pi\tau f''_w(\lambda^*)}}\,
R_{5}\biggl(\sqrt{\tau \bigl[h_\text{c}(w)-h_\text{s}(w)\bigr]}, 
\sqrt{\tau \bigl[h_\text{a}(w)-h_\text{s}(w)\bigr]}, 
\sqrt{\tau \bigl[h_\text{d}(w)-h_\text{s}(w)\bigr]}\biggr),
\label{P(W)-right}
\end{align}
where $R_3(z)$ is given by
\eref{R2tilde}, and 
\begin{equation}
R_{5}(z_1,z_2,z_3) = \sqrt\frac{z_1 z_2 z_3}{\pi}
\int_{0}^\infty  du\, e^{-u^2}
\biggl[
  \frac{1}{\sqrt{z_1+iu}\sqrt{z_2+i u}\sqrt{z_3+iu}}  
-  \frac{1}{\sqrt{z_1-iu}\sqrt{z_2-iu}\sqrt{z_3-iu}}
\biggr]i. 
\end{equation}

\subsubsection{$w^*_\text{a} < w < w^*_\text{c}$}

 For $w^*_\text{a} < w < w^*_\text{c}$, the saddle point lies between
$\lambda_\text{c}$ and $\lambda_\text{a}$. Therefore, the contour of
integration can be deformed through the saddle point without crossing
any singularity, as shown in \fref{steepest-descent
contour3}~(b). Now, to compute the saddle point contribution one can
follow the methods of \aref{sec: branch point outside}, while taking
into account of both the singularities $\lambda_\text{a}$ and
$\lambda_\text{c}$. The calculation yields
\begin{equation}
P(W_{\tau}=w\tau)\approx
\frac{g(\lambda^*) e^{\tau h_\text{s}(w)}}
{\sqrt{2\pi\tau f''_w(\lambda^*)}}R_{6}\biggl(\sqrt{\tau \bigl[h_\text{c}(w)-h_\text{s}(w)\bigr]}, 
\sqrt{\tau \bigl[h_\text{a}(w)-h_\text{s}(w)\bigr]}, 
\sqrt{\tau \bigl[h_\text{d}(w)-h_\text{s}(w)\bigr]}\biggr),
\label{P(W)-mid}
\end{equation}
where 
\begin{equation}
R_6(z_1,z_2,z_3)=\sqrt\frac{z_1 z_2 z_3}{\pi}
\int_{-\infty}^\infty \frac{e^{-u^2}\, du}{\sqrt{z_1+i u}\sqrt{z_2-i u}
\sqrt{z_3-i u}}.
\label{R6}
\end{equation}
As $w\rightarrow w^*_\text{c}$, the first term of \eref{P(W)-right},
coming from the integral along the branch cut, goes to zero. On the
other hand, it can be shown that $R_5(z_1\rightarrow 0, z_2, z_3)=
R_6(z_1\rightarrow 0, z_2, z_3)$. Therefore, Eqs.~\eqref{P(W)-right}
and \eqref{P(W)-mid} approach the same limiting form as $w\rightarrow
w^*_\text{c}$ from the two sides.

\subsubsection{$w^*_\text{d} < w < w^*_\text{a}$}

 For $w^*_\text{d} < w < w^*_\text{a}$, the saddle point lies between
$\lambda_\text{a}$ and $\lambda_\text{d}$.  Therefore, the deformed
contour is as shown in \fref{steepest-descent contour3}~(c).
Combining the contributions from the branch point $\lambda_\text{a}$
and the saddle point, we get
\begin{align}
P(W_{\tau}=w\tau)\approx  &
\frac{\widetilde{g}(\lambda_\text{a})\, e^{\tau h_\text{a}(w)}}
{\sqrt{\pi\tau|f'_w(\lambda_\text{a})|}} 
R_7\biggl(\sqrt{\tau \bigl[h_\text{a}(w)-h_\text{s}(w)\bigr]},
\sqrt{\tau \bigl[h_\text{d}(w)-h_\text{s}(w)\bigr]}\biggr)\notag\\
+&\frac{|g(\lambda^*)|\, e^{\tau h_\text{s}(w)}}
{\sqrt{2\pi\tau f''_w(\lambda^*)}} 
R_8\biggl(\sqrt{\tau \bigl[h_\text{c}(w)-h_\text{s}(w)\bigr]},
\sqrt{\tau \bigl[h_\text{a}(w)-h_\text{s}(w)\bigr]},
\sqrt{\tau \bigl[h_\text{d}(w)-h_\text{s}(w)\bigr]}\biggr),
\label{P(W)-mid2}
\end{align}
where
\begin{align}
\label{R7}
R_7(z_1,z_2)=&\sqrt\frac{2z_1(z_1+z_2)}{\pi}
\int_0^{z_1} \frac{e^{-2z_1u + u^2}}{\sqrt{u}\sqrt{z_1+z_2-u}}\, du,
\\
\label{R8}
\text{and}\quad
R_8(z_1,z_2,z_3) 
= &\sqrt\frac{z_1 z_2 z_3}{\pi}
\int_{0}^\infty du\,e^{-u^2}
\left[
  \frac{1}{\sqrt{z_1+iu}\sqrt{z_2+iu}\sqrt{z_3-iu}} - 
  \frac{1}{\sqrt{z_1-iu}\sqrt{z_2-iu}\sqrt{z_3+iu}}
\right]i .
\end{align}
As $w\rightarrow w^*_\text{a}$, the first term of \eref{P(W)-mid2},
coming from the integral along the branch cut, goes to zero. On the
other hand, it can be shown that $R_6(z_1, z_2\rightarrow 0, z_3)=
R_8(z_1, z_2\rightarrow 0, z_3)$. Therefore, Eqs.~\eqref{P(W)-mid}
and \eqref{P(W)-mid2} approach the same limiting form as $w\rightarrow
w^*_\text{a}$ from the two sides.

\subsubsection{$w < w^*_\text{d}$}
 
Finally, for $w < w^*_\text{d}$, the saddle point lies between
$\lambda_\text{d}$ and $\lambda_+$. In this case, the integral along
the branch cut can be divided into two parts: one, from
$\lambda_\text{a}$ to $\lambda_\text{d}$ and another from
$\lambda_\text{d}$ to $\lambda^*$.  Between $\lambda_\text{d}$ and
$\lambda^*$, the the integral above the branch cut exactly cancels the
integral below the branch cut. Therefore, the net contribution is the
sum of the contributions coming from the integral around the branch
cut between $\lambda_\text{a}$ and $\lambda_\text{d}$, and the
contribution of the integral along the contour ($C_1$ and $C_4$)
through the saddle point, for which the calculation is similar to the
one given in \aref{sec: branch point outside}. Therefore, we get
\begin{align}
P(W_{\tau}=w\tau) \approx &
\frac{\widetilde{g}(\lambda_\text{a})\, e^{\tau h_\text{a}(w)}}
{\sqrt{\pi\tau|f'_w(\lambda_\text{a})|}} 
R_{9}\biggl(\sqrt{\tau \bigl[h_\text{a}(w)-h_\text{s}(w)\bigr]},
\sqrt{\tau \bigl[h_\text{d}(w)-h_\text{s}(w)\bigr]}\biggr)\notag\\
-&\frac{|g(\lambda^*)|\, e^{\tau h_\text{s}(w)}}
{\sqrt{2\pi\tau f''_w(\lambda^*)}} 
R_{10}\biggl(\sqrt{\tau \bigl[h_\text{c}(w)-h_\text{s}(w)\bigr]},
\sqrt{\tau \bigl[h_\text{a}(w)-h_\text{s}(w)\bigr]},
\sqrt{\tau \bigl[h_\text{d}(w)-h_\text{s}(w)\bigr]}\biggr),
\label{P(W)-left}
\end{align}
where
\end{widetext}
\begin{align}
\label{R9}
&R_{9}(z_1,z_2)=\sqrt\frac{2z_1(z_1-z_2)}{\pi}
\int_0^{z_1-z_2} \frac{e^{-2z_1u + u^2}\, du}{\sqrt{u}\sqrt{z_1-z_2-u}},
\\
\label{R10}
&R_{10}(z_1,z_2,z_3)=\sqrt\frac{z_1 z_2 z_3}{\pi}
\int_{-\infty}^\infty \frac{e^{-u^2}\, du}{\sqrt{z_1+i u}\sqrt{z_2+i u}
\sqrt{z_3+i u}}.
\end{align}
It is evident from the above equations that $R_7(z_1,0)=R_9(z_1,0)$.
Moreover, it can be shown that $R_{10}(z_1, z_2, z_3\rightarrow 0)= -
R_8(z_1, z_2, z_3\rightarrow 0)$. Therefore, Eqs.~\eqref{P(W)-mid2}
and \eqref{P(W)-left} approach the same limiting form as $w\rightarrow
w^*_\text{d}$ from the two sides.

\subsection{The case of four singularities}
\label{four-singularities}

Finally, we consider the case $\delta>1$ and $\theta < \theta_{c_2}$,
in which case $g(\lambda)$ has four singularities (see \fref{phase2})
at $\lambda_\text{a}$, $\lambda_\text{b}$, $\lambda_\text{c}$ and
$\lambda_\text{d}$ given by Eqs. \eqref{lambda-a}--\eqref{lambda-d}
respectively; where $\lambda_- < \lambda_\text{b} < \lambda_\text{c} <
0 < \lambda_\text{a} < \lambda_\text{d}<\lambda_+$.  Therefore, and
$g(\lambda)$ can be written as
\begin{equation}
g(\lambda)=\frac{g_4(\lambda)}{
\sqrt{\lambda-\lambda_\text{b}}
\sqrt{\lambda-\lambda_\text{c}}
\sqrt{\lambda_\text{a}-\lambda}
\sqrt{\lambda_\text{d}-\lambda}},
\end{equation}
where $g_4(\lambda)$ is the analytical factor of $g(\lambda)$.

Now as $w$ varies from $+\infty$ to $-\infty$, the saddle point hits
the branch points, $\lambda^*(w^*_i)=\lambda_i$ with $i\in\{\text{b,
c, a, d}\}$, at specific values of $w$ given by \eref{wstar} and
$+ \infty > w^*_\text{b} > w^*_\text{c} > w^*_\text{a} > w^*_\text{d}
>-\infty$. It is straightforward to generalize the above results to
this case of four singularities. Therefore, we only give the results
below, without repeating the details.

\begin{widetext}

\subsubsection{$w > w^*_\text{b}$}

 For $w > w^*_\text{b}$, the saddle point lies between $\lambda_-$ and
$\lambda_\text{b}$, and
\begin{align}
P(W_{\tau}=w\tau)\approx &
\frac{\widetilde{g}(\lambda_\text{c})\, e^{\tau
h_\text{c}(w)}}{\sqrt{\pi\tau|f'_w(\lambda_\text{c})|}} 
R_{9}\biggl(\sqrt{\tau \bigl[h_\text{c}(w)-h_\text{s}(w)\bigr]},
\sqrt{\tau \bigl[h_\text{b}(w)-h_\text{s}(w)\bigr]}\biggr)\notag\\
-&\frac{|g(\lambda^*)|\, e^{\tau h_\text{s}(w)}}
{\sqrt{2\pi\tau f''_w(\lambda^*)}} 
R_{11}\biggl(\sqrt{\tau \bigl[h_\text{b}(w)-h_\text{s}(w)\bigr]},
\sqrt{\tau \bigl[h_\text{c}(w)-h_\text{s}(w)\bigr]},
\sqrt{\tau \bigl[h_\text{a}(w)-h_\text{s}(w)\bigr]},
\sqrt{\tau \bigl[h_\text{d}(w)-h_\text{s}(w)\bigr]}\biggr),
\label{PDF-1}
\end{align}
where $R_{9}(z_1,z_2)$ is given by \eref{R9} and
\begin{equation}
\label{R11}
R_{11}(z_1,z_2,z_3,z_4)=\sqrt\frac{z_1 z_2 z_3 z_4}{\pi}
\int_{-\infty}^\infty \frac{e^{-u^2}\, du}{\sqrt{z_1-i u}\sqrt{z_2-i u}
\sqrt{z_3-i u}\sqrt{z_4-i u}}=
\sqrt\frac{z_1 z_2 z_3 z_4}{\pi}
\int_{-\infty}^\infty \frac{e^{-u^2}\, du}{\sqrt{z_1+i u}\sqrt{z_2+i u}
\sqrt{z_3+i u}\sqrt{z_4+i u}}.
\end{equation}

\subsubsection{$w^*_\text{c} <w < w^*_\text{b}$}

For $w^*_\text{c} <w < w^*_\text{b}$, the saddle point lies between
$\lambda_\text{b}$ and $\lambda_\text{c}$, and
\begin{align}
P(W_{\tau}=w\tau)\approx  
&
\frac{\widetilde{g}(\lambda_\text{c})\,e^{\tau
h_\text{c}(w)}}{\sqrt{\pi\tau|f'_w(\lambda_\text{c})|}} 
R_7\biggl(\sqrt{\tau \bigl[h_\text{c}(w)-h_\text{s}(w)\bigr]},
\sqrt{\tau \bigl[h_\text{b}(w)-h_\text{s}(w)\bigr]}\biggr)\notag\\
+&\frac{|g(\lambda^*)|\, e^{\tau h_\text{s}(w)}}
{\sqrt{2\pi\tau f''_w(\lambda^*)}} 
R_{12}\biggl(\sqrt{\tau \bigl[h_\text{b}(w)-h_\text{s}(w)\bigr]},
\sqrt{\tau \bigl[h_\text{c}(w)-h_\text{s}(w)\bigr]},
\sqrt{\tau \bigl[h_\text{a}(w)-h_\text{s}(w)\bigr]},
\sqrt{\tau \bigl[h_\text{d}(w)-h_\text{s}(w)\bigr]}\biggr),
\label{PDF-2}
\end{align}
where $R_7(z_1,z_2)$ is given by \eref{R7} and
\begin{equation}
R_{12}(z_1,z_2,z_3,z_4) 
= \sqrt\frac{z_1 z_2 z_3 z_4}{\pi}
\int_{0}^\infty du\,e^{-u^2}
\left[
  \frac{1}{\sqrt{z_1-iu}\sqrt{z_2+iu}\sqrt{z_3+iu}\sqrt{z_4+iu}} -
  \frac{1}{\sqrt{z_1+iu}\sqrt{z_2-iu}\sqrt{z_3-iu}\sqrt{z_4-iu}} 
\right]i .
\end{equation}

\subsubsection{$w^*_\text{a} <w < w^*_\text{c}$}

For $w^*_\text{a} <w < w^*_\text{c}$, the saddle point lies between
$\lambda_\text{c}$ and $\lambda_\text{a}$, and the PDF is given by
\begin{equation}
P(W_{\tau}=w\tau)\approx  
\frac{g(\lambda^*)\, e^{\tau h_\text{s}(w)}}
{\sqrt{2\pi\tau f''_w(\lambda^*)}} 
R_{13}\biggl(\sqrt{\tau \bigl[h_\text{b}(w)-h_\text{s}(w)\bigr]},
\sqrt{\tau \bigl[h_\text{c}(w)-h_\text{s}(w)\bigr]},
\sqrt{\tau \bigl[h_\text{a}(w)-h_\text{s}(w)\bigr]},
\sqrt{\tau \bigl[h_\text{d}(w)-h_\text{s}(w)\bigr]}\biggr),
\label{PDF-3}
\end{equation}
where
\begin{equation}
R_{13}(z_1,z_2,z_3,z_4)=\sqrt\frac{z_1 z_2 z_3 z_4}{\pi}
\int_{-\infty}^\infty \frac{e^{-u^2}\, du}{\sqrt{z_1+i u}\sqrt{z_2+i
u}
\sqrt{z_3-i u} \sqrt{z_4-i u}}.
\end{equation}

\subsubsection{$w^*_\text{d} <w < w^*_\text{a}$}

For $w^*_\text{d} <w < w^*_\text{a}$, the saddle point lies between
$\lambda_\text{a}$ and $\lambda_\text{d}$, and
\begin{align}
P(W_{\tau}=w\tau)\approx  
&
\frac{\widetilde{g}(\lambda_\text{a})\, e^{\tau
h_\text{a}(w)}}{\sqrt{\pi\tau|f'_w(\lambda_\text{a})|}} 
R_7\biggl(\sqrt{\tau \bigl[h_\text{a}(w)-h_\text{s}(w)\bigr]},
\sqrt{\tau \bigl[h_\text{d}(w)-h_\text{s}(w)\bigr]}\biggr)\notag\\
+&\frac{|g(\lambda^*)|\, e^{\tau h_\text{s}(w)}}
{\sqrt{2\pi\tau f''_w(\lambda^*)}} 
R_{14}\biggl(\sqrt{\tau \bigl[h_\text{b}(w)-h_\text{s}(w)\bigr]},
\sqrt{\tau \bigl[h_\text{c}(w)-h_\text{s}(w)\bigr]},
\sqrt{\tau \bigl[h_\text{a}(w)-h_\text{s}(w)\bigr]},
\sqrt{\tau \bigl[h_\text{d}(w)-h_\text{s}(w)\bigr]}\biggr),
\label{PDF-4}
\end{align}
where $R_7(z_1,z_2)$ is given by \eref{R7} and
\begin{equation}
R_{14}(z_1,z_2,z_3,z_4) 
= \sqrt\frac{z_1 z_2 z_3 z_4}{\pi}
\int_{0}^\infty du\,e^{-u^2}
\left[
  \frac{1}{\sqrt{z_1+iu}\sqrt{z_2+iu}\sqrt{z_3+iu}\sqrt{z_4-iu}} - 
  \frac{1}{\sqrt{z_1-iu}\sqrt{z_2-iu}\sqrt{z_3-iu}\sqrt{z_4+iu}}
\right]i .
\end{equation}

\subsubsection{$w < w^*_\text{d}$}

 Finally, for $w < w^*_\text{d}$, the saddle point lies between
$\lambda_\text{d}$ and $\lambda_+$, and
\begin{align}
P(W_{\tau}=w\tau)\approx  
&
\frac{\widetilde{g}(\lambda_\text{a})\, e^{\tau
h_\text{a}(w)}}{\sqrt{\pi\tau|f'_w(\lambda_\text{a})|}} 
R_{9}\biggl(\sqrt{\tau \bigl[h_\text{a}(w)-h_\text{s}(w)\bigr]},
\sqrt{\tau \bigl[h_\text{d}(w)-h_\text{s}(w)\bigr]}\biggr)\notag\\
-&\frac{|g(\lambda^*)|\, e^{\tau h_\text{s}(w)}}
{\sqrt{2\pi\tau f''_w(\lambda^*)}} 
R_{11}\biggl(\sqrt{\tau \bigl[h_\text{b}(w)-h_\text{s}(w)\bigr]},
\sqrt{\tau \bigl[h_\text{c}(w)-h_\text{s}(w)\bigr]},
\sqrt{\tau \bigl[h_\text{a}(w)-h_\text{s}(w)\bigr]},
\sqrt{\tau \bigl[h_\text{d}(w)-h_\text{s}(w)\bigr]}\biggr).
\label{PDF-5}
\end{align}
where $R_9(z_1,z_2)$ and $R_{11}(z_1,z_2,z_3,z_4)$ are given by
Eqs.~\eqref{R9} and \eqref{R11}, respectively.

\end{widetext}

It can be shown that, when $w\rightarrow w_i^*$ with $i\in \{\text{a},
\text{b}, \text{c}, \text{d}\}$ from the two sides of $w^*_i$, the respective
expressions of the PDFs, i.e, Eqs.~\eqref{PDF-1} and \eqref{PDF-2},
Eqs.~\eqref{PDF-2} and \eqref{PDF-3}, Eqs.~\eqref{PDF-3}
and \eqref{PDF-4}, and Eqs.~\eqref{PDF-4} and \eqref{PDF-5},
respectively, approach the same limiting form.

\subsubsection{Numerical simulation}

We now compare the analytical results obtained in this section with
numerical simulation. We consider $\delta=5$, for which we have
$\theta_{c_2}=0.18$.  Therefore, $g(\lambda)$ has three singularities
for $\theta > \theta_{c_2}$, whereas for $\theta <\theta_{c_2}$ it has
four singularities.

\begin{figure}
\includegraphics[width=\hsize]{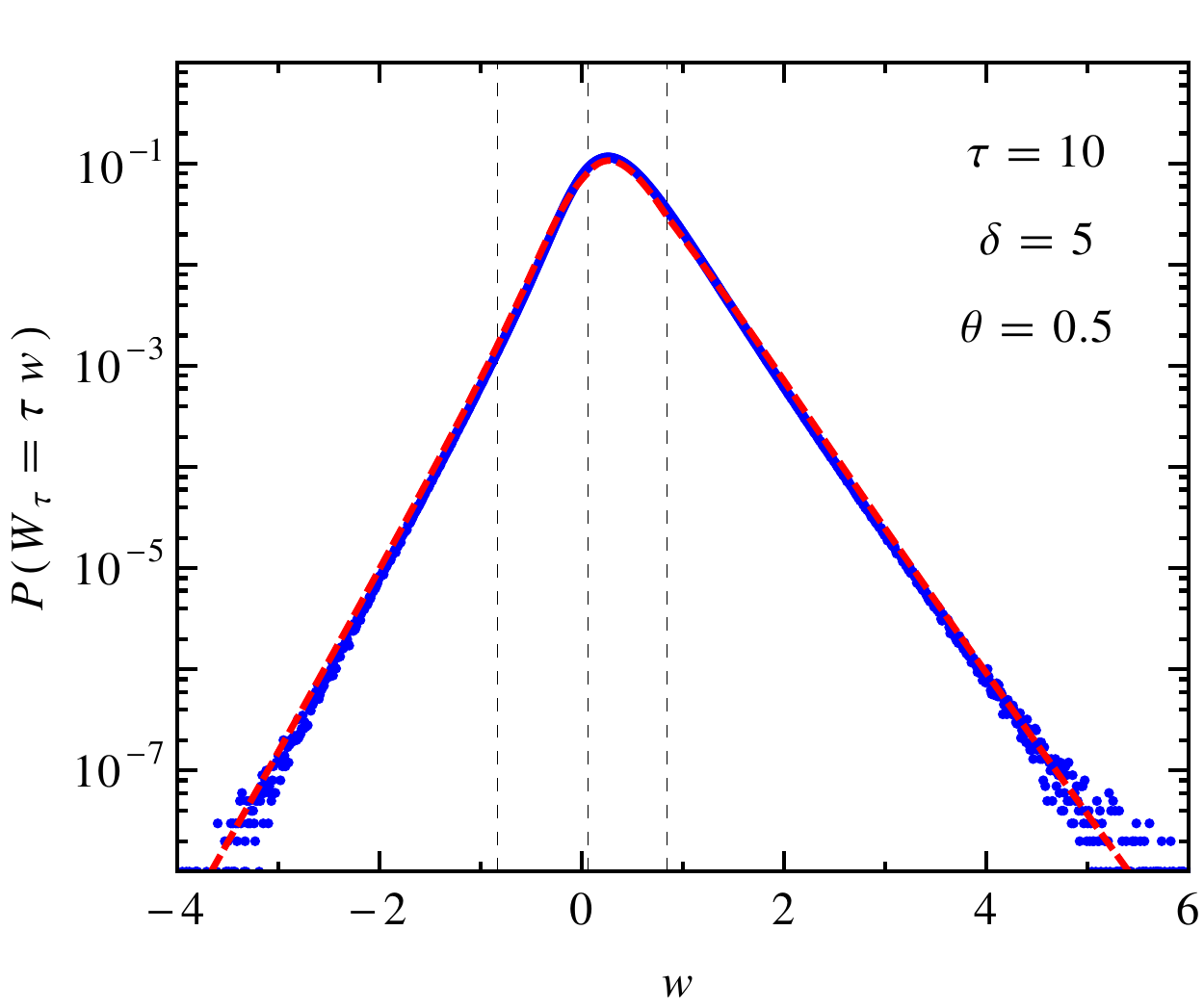}
\caption{\label{pdf3} (Color online). $P(W_\tau)$ against the scaled
variable $w=W_\tau/\tau$ for $\tau=10$, $\tau_c=1$, $\delta=5$, and
 $\theta=0.5$. The points (blue) are obtained from numerical
 simulation, and the dashed solid line (red) plots the analytical
 asymptotic forms given in the text.  The vertical dashed lines mark
 the positions $w^*_\text{c}=0.8398\dots$, $w^*_\text{a}=0.06269\dots$
 and $w^*_\text{d}=-0.8398\dots$.}
\end{figure}

For $\theta=0.5$ the three singularities are located at
$\lambda_\text{c}=-0.3062\dots$, $\lambda_\text{a}=0.3958\dots$ and
$\lambda_\text{d}=1.3062\dots$, whereas $\lambda_-=-0.4848\dots$ and
$\lambda_+=1.4848\dots$. \Fref{pdf3} compares numerical simulation for
this case with analytical results obtained above for the case of the
three singularities.

\begin{figure}
\includegraphics[width=\hsize]{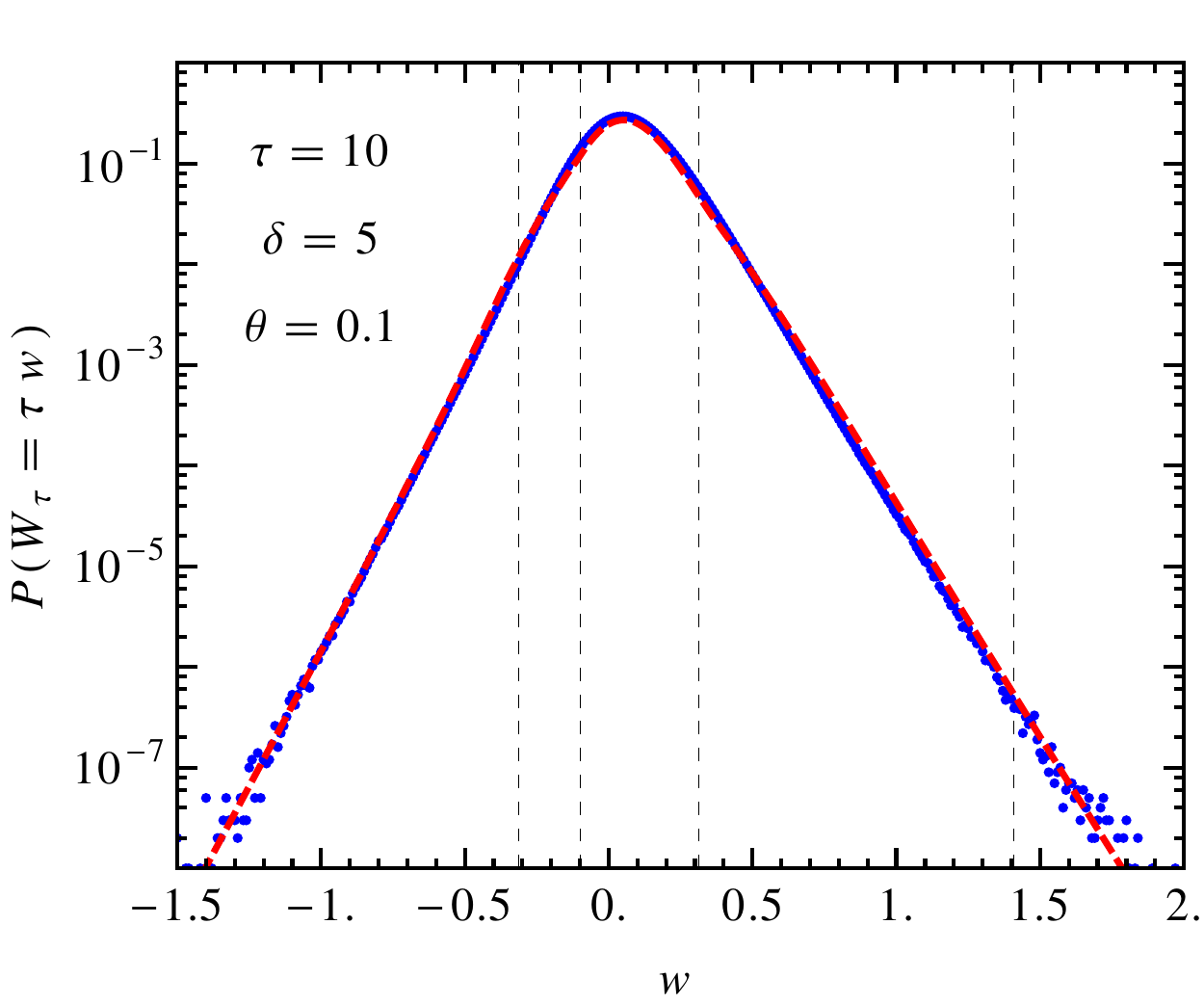}
\caption{\label{pdf4} (Color online). $P(W_\tau)$ against the scaled
variable $w=W_\tau/\tau$ for $\tau=10$, $\tau_c=1$, $\delta=5$ and
 $\theta=0.1$. The points (blue) are obtained from numerical
 simulation, and the dashed solid line (red) plots the analytical
 asymptotic forms given in the text.  The vertical dashed lines mark
 the positions $w^*_\text{b}=1.4062\dots$, $w^*_\text{c}=0.3125$,
 $w^*_\text{a}=-0.0976\dots$ and $w^*_\text{d}=-0.3125$.}
\end{figure}

On the other hand, for $\theta=0.1$, the four singularities of
$g(\lambda)$ are located at at $\lambda_\text{b}=-10/7$,
$\lambda_\text{c}=-1$, $\lambda_\text{a}=13/11$, and
$\lambda_\text{d}=2$. Moreover, $\lambda_-=-1.4621\dotsc$ and
$\lambda_+=2.4621\dotsc$.  \Fref{pdf4} compares numerical simulation
for this case with analytical results obtained above for the case of
the four singularities.

\section{Large Deviation Function and Fluctuation Theorem}
\label{sec:LDF}

The large deviation function is defined by
\begin{equation}
h(w)=\lim_{\tau\rightarrow\infty}\frac{1}{\tau}\ln P(W_\tau=w\tau).
\label{large-deviation function}
\end{equation}
In other words, the large deviation form of the PDF refers to the
ultimate asymptotic form $P(W_\tau=w\tau)\sim e^{\tau h(w)}$ while
ignoring the subleading corrections. Apart from being an interesting
quantity on its own, the large deviation functions have found
importance recently in the context of the fluctuation theorem. The
latter refers to the relation
\begin{equation}
\lim_{\tau\rightarrow\infty}\frac{1}{\tau}\ln 
\left[
\frac{P(W_\tau=+w\tau)}
{P(W_\tau=-w\tau)}\right]=w.
\label{fluctuation theorem}
\end{equation}
When the above relation is valid, the large deviation function
evidently satisfies the symmetry relation
\begin{equation}
h(w)-h(-w)=w.
\label{LDF symmetry}
\end{equation}

Now, as we have seen in the above sections, when $g(\lambda)$ is
analytic between the origin and the saddle point, the dominant
contribution to $P(W_\tau)$ comes from the saddle point as given
by \eref{saddle-point approximation}. On the other hand, when there
are singularities between the origin and the saddle point, the most
dominant contribution to $P(W_\tau)$ comes from the singularity
closest to the origin (farthest from the saddle point) and lies
between the origin and the saddle point. This is because, evidently
$-\nu(\lambda)$, and hence the function $f_w(\lambda)$, is convex on
the interval $[\lambda_-,\lambda_+]$ and $f_w(\lambda)$ is minimum at
the saddle point $\lambda^*$ along the real-$\lambda$ line.

Consequently, for the case $\delta <1$ and $\theta<\theta_{c_1}$,
where $g(\lambda)$ is analytic on the interval
$(\lambda_-,\lambda_+)$, the large deviation function is
$h(w)=h_s(w)$, given by \eref{hs(w)}. In this case, $h(w)$ satisfies
the above symmetry relation \eqref{LDF symmetry}, and therefore, the
fluctuation theorem is valid.  On the other hand, for $\delta<1$ and
$\theta>\theta_{c_1}$, where $g(\lambda)$ has one singularity at
$\lambda_\text{a}$, (also for $\delta=1$ and all values of $\theta$,
where only the singularity at $\lambda_\text{a}$ is relevant), one has
\begin{equation}
h(w)=\begin{cases}
h_s(w) &\text{for} ~w> w^*_\text{a},\\
h_\text{a}(w)  &\text{for} ~w < w^*_\text{a}.
\end{cases}
\end{equation}
Therefore, it is only when $w^*_\text{a} <0$ (e.g., when $\theta <4$
for the $\delta=1$ case), the symmetry relation \eref{LDF symmetry}
(and hence the fluctuation theorem) is satisfied only in the specific
range $w^*_\text{a} < w < -w^*_\text{a}$. Otherwise it is not
satisfied.

For the case $\delta>1$, although there are either three or four
singularities depending on whether $\theta >\theta_{c_2}$ or $\theta
<\theta_{c_2}$, the singularities closest to the origin (one on each
side), namely $\lambda_\text{c}$ and $\lambda_\text{a}$ are common in
both cases. Therefore, for both cases, the large deviation function is
given by
\begin{equation}
h(w)=\begin{cases}
h_\text{c}(w)  &\text{for} ~w > w^*_\text{c},\\
h_s(w) &\text{for} ~w^*_\text{a} < w < w^*_\text{c},\\
h_\text{a}(w)  &\text{for} ~w < w^*_\text{a}.
\end{cases}
\end{equation}
Since $\lambda_\text{c} < 0$, it is evident from \eref{wstar} that
$w^*_\text{c} >0$. Therefore again, it is only when $w^*_\text{a} <0$
(e.g., when $\theta <0.365\dots$ for the $\delta=5$ case), the
symmetry relation \eref{LDF symmetry} (and hence the fluctuation
theorem) is satisfied only in the specific range
$\max(w^*_\text{a},-w^*_\text{c}) < w
< \min(-w^*_\text{a},w^*_\text{c})$.

Therefore, for any $\delta$, there exists a $\theta_c$, given by
$w^*_\text{a}=0$ (equivalently $\lambda_\text{a}=1/2$) as
\begin{equation}
\theta_c(\delta)=\frac{3+2 \delta +3 \delta ^2
+(1-\delta) \sqrt{9+14 \delta +9 \delta ^2}}{2 \delta ^2},
\label{theta_c} 
\end{equation}
and the fluctuation theorem is not valid for $\theta >\theta_c$.  The
$\theta=\theta_c(\delta)$ line corresponds to the
$\alpha=\alpha_c(\delta)$ line in the $(\alpha, \delta)$ plane where
\begin{equation}
\alpha_c(\delta)=\frac{3+2 \delta +3 \delta ^2
+(1-\delta) \sqrt{9+14 \delta +9 \delta ^2}}{2 (1+\delta)}.
\label{alpha_c} 
\end{equation}

\begin{figure*}
\includegraphics[width=.48\hsize]{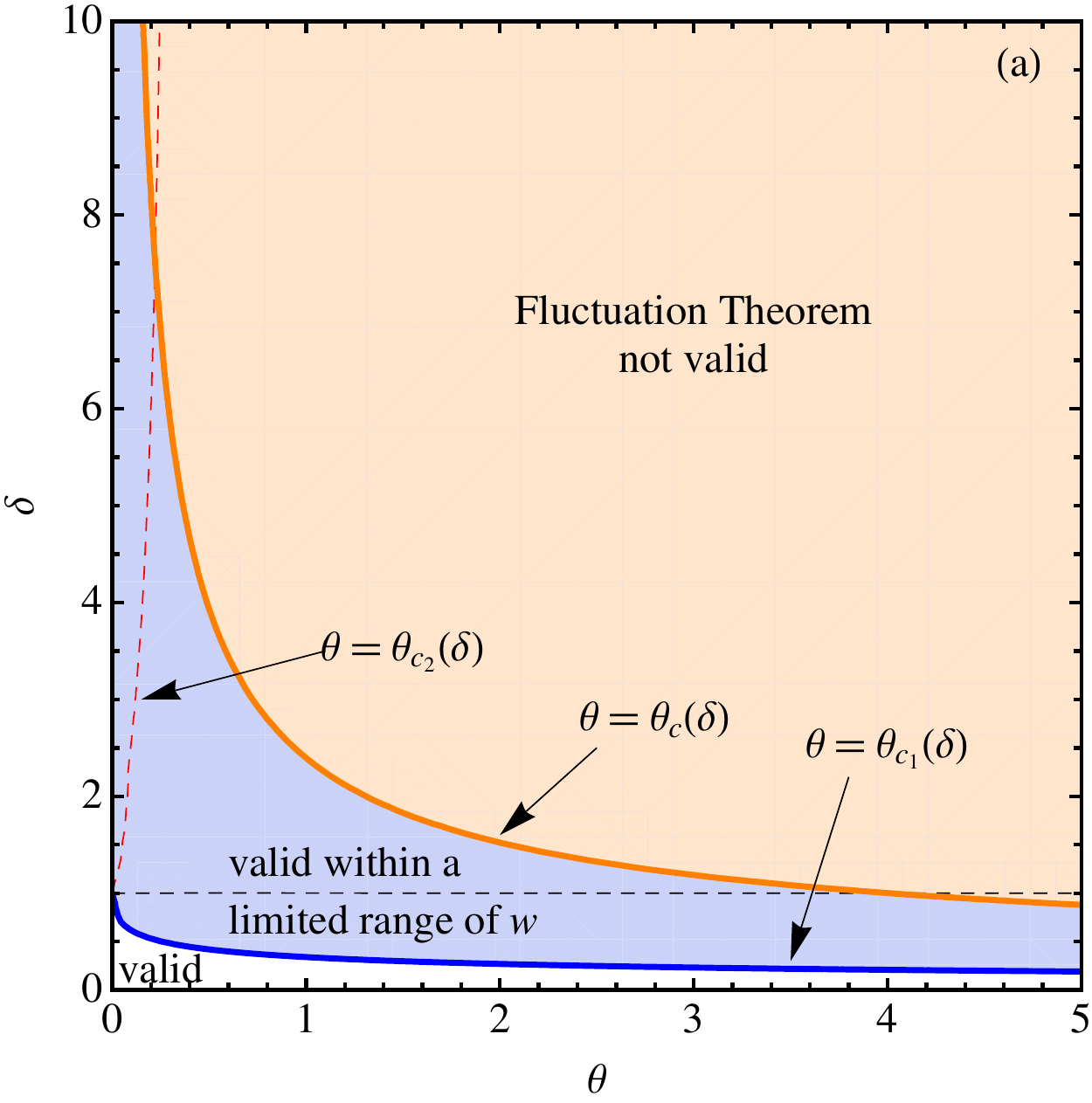}
\includegraphics[width=.475\hsize]{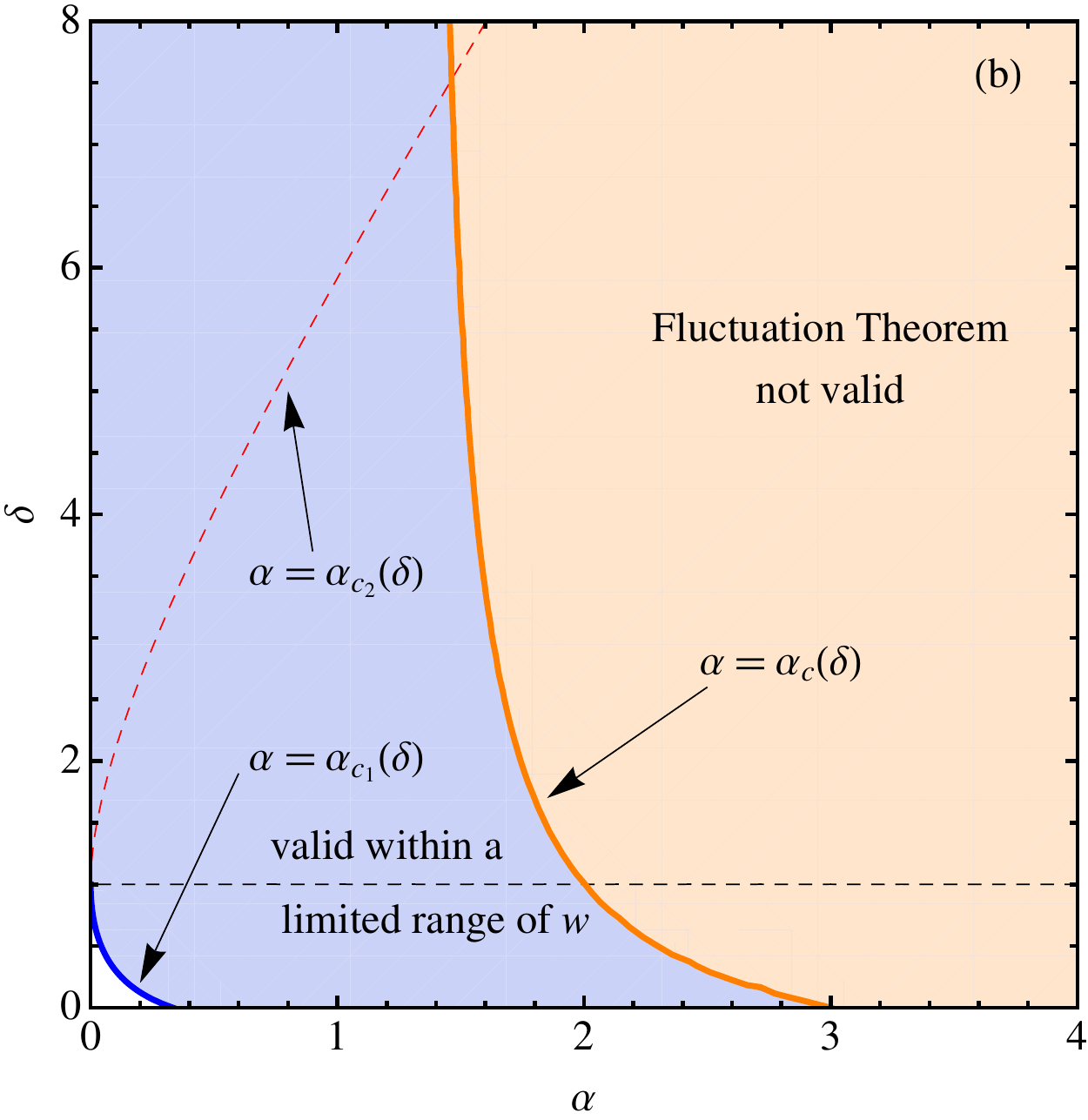}
\caption{\label{phase3} (Color online). The phase diagrams in the  parameters
space showing the state of the validity of the fluctuation
theorem. Apart from the $\theta=\theta_c$ and $\alpha=\alpha_c$
(orange) lines, the other lines are the same as in \fref{phase2}. In
the (light orange) regions above the (orange) lines $\theta=\theta_c$
in (a) and $\alpha=\alpha_c$ in (b), the fluctuation theorem is not
valid at all, whereas it is always valid in the white regions below
the (blue) lines $\theta=\theta_{c_1}$ in (a) and
$\alpha=\alpha_{c_1}$ in (b). In the intermediate (light blue) region,
the fluctuation theorem is valid only within a limited range of $w$
given by $w^*_\text{a} < w < -w^*_\text{a}$ for $\delta<1$ and
$\max(w^*_\text{a},-w^*_\text{c}) < w
< \min(-w^*_\text{a},w^*_\text{c})$ for $\delta >1$. For $\delta=0$,
we have $\alpha_{c_1}=1/3$ and $\alpha_c=3$.}
\end{figure*}

\Fref{phase3} summarizes the state of validity of the fluctuation
theorem in the $\delta, \theta$ and $\alpha, \delta$ parameter spaces.

\section{Summary}
\label{sec:summary}

Let us now summarize the main contents of the paper.  We have obtained
analytical results for a system studied recently
experimentally~\cite{Ciliberto:10}. The experimental system consists
of a colloidal particle in water and confined in an optical trap which
is modulated according to an Ornstein-Uhlenbeck process. This system
is described by a set of coupled Langevin equations. We have computed
the PDF of the work done by the modulating trap on the Brownian
particle in a given time $\tau$, for large $\tau$. The moment
generating function of the work has the $\bigl \langle e^{-\lambda
W_\tau}\bigr\rangle
\approx g(\lambda)\,   e^{\tau\mu(\lambda)}$ for  large
$\tau$. Inverting this, we obtain the PDF of the work within the
saddle point approximation.  

The results can be described in terms of two independent parameters:
(1) the parameter $\theta$ that quantifies the relative strength of
the external noise that generates the Ornstein-Uhlenbeck process for
the trap modulation, with respect to the thermal fluctuations, and (2)
the ratio $\delta=\tau_0/\tau_\gamma$ of the correlation time of the
trap modulation to the viscous relaxation time of the particle in the
trap without any modulation. We find that the cumulant generating
function $\mu(\lambda)$ is analytic in a (real) interval
$(\lambda_-,\lambda_+)$ and the saddle point lies within this interval
--- here $\lambda_\pm$ depends on the values of $\theta, \delta$.  On
the other hand, depending on the values of the pair
$(\theta, \delta)$, the function $g(\lambda)$ behaves differently.
For $\delta < 1$, there exists a value $\theta_{c_1} (\delta)$ such
that $g(\lambda)$ is analytic in the interval $(\lambda_-,\lambda_+)$
for $\theta < \theta_{c_1}$ whereas it has a branch point for $\theta
> \theta_{c_1}$. For $\delta >1$, there again exists a
$\theta_{c_2}(\delta)$, and $g(\lambda)$ has either three or four
branch points depending on whether $\theta >\theta_{c_2}$ or $\theta
<\theta_{c_2}$.  For $\delta=1$, there are three branch points of
which two coincide with $\lambda_\pm$. We have done the analysis in
each of these regions and obtained the asymptotic form of the PDF
accordingly. We have compared our analytical results with simulation
results on this system and found very good agreement between the two.

The calculation also gives the large deviation function as a
by-product, using which we check the validity of the so-called
fluctuation theorem for this context. We find that in the region
$\delta <1, \theta <\theta_{c_1}$, it is always valid. Outside this
parameter region, there exists a $\theta_c(\delta)$ and the
fluctuation theorem is valid for a limited range of $w$ around zero
when $\theta<\theta_c$. For $\theta>\theta_c$, the fluctuation theorem
is not valid at all (see \fref{phase3}).

Finally, we would like to point out that it is very easy to generalize
the analysis of the paper to the case of a Brownian particle subjected
to any exponentially correlated external random force. Moreover, the
asymptotic analysis (steepest descent method with branch points.)
carried out in this paper should be applicable for finding asymptotic
approximations of similar integrals in general.

\begin{acknowledgements}
SS acknowledges the support of the Indo-French Centre for the
Promotion of Advanced Research (IFCPAR/CEFIPRA) under Project
no. 4604-3.
\end{acknowledgements}

\appendix

\section{Calculation of the Moment Generating Function}
\label{MGF calculation}

The evolution equations \eqref{Langevin-1} and \eqref{Langevin-2} can
be presented in the matrix form
\begin{eqnarray} 	 
\frac{dU}{dt}=-\frac{1}{\tau_{0}}{A} U + \eta(t),
\label{Langevin}
\end{eqnarray}
where $U=(x,y)^T$ and $\eta=(\xi,\zeta)^T$ are column vectors and
 ${A}$ is a $2\times2$ matrix given by
\begin{equation}
{A}=\begin{pmatrix}
\delta & -\delta \\
0 & 1
\end{pmatrix}.
\end{equation}

Using the integral representation of $\delta$-function,
the \emph{restricted} moment generating function defined
by \eref{restricted GF}
\begin{equation}
Z(\lambda,U,\tau|U_{0})=\int \frac{d^{2}\sigma}{(2\pi)^{2}}
e^{i\sigma^{T}U} 
\bigl\langle e^{-\lambda
W_{\tau}-i\sigma^{T}U(\tau)}\bigr\rangle_{U_{0}}, 
\label{R-GF}
\end{equation}
where $\sigma^{T}=(\sigma_{1},\sigma_{2})$.

Substituting $dx/dt$ form \eref{Langevin-1} in \eref{work} we get
\begin{equation}
W_{\tau}=\int_{0}^{\tau}dt\, \left[-\frac{\delta^{2}}{ \tau_{0}^{2}D} y(x-y)
+ \frac{\delta}{ \tau_{0}D}  y \xi\right],
\label{work2}
\end{equation}
which is useful to rewrite as
\begin{equation}
W_{\tau}
=\frac{1}{2} \int_{0}^{\tau} dt
\left[
\frac{1}{\tau_{0}^2} U^T{A}_{1}U +
\frac{1}{\tau_{0}}
\bigl( U^T{A}_{2}^T\eta+\eta^T{A}_{2}U\bigr)
\right], 
\label{work3}
\end{equation}
where
\begin{equation}
{A}_{1}=\frac{\delta^{2}}{D}
\begin{pmatrix}
0 & -1 \\
-1 & 2\\
\end{pmatrix}
\quad\text{and}\quad
{A}_{2}=\frac{\delta}{D}
\begin{pmatrix}
0 & 1 \\
0 & 0\\
\end{pmatrix}.
\end{equation}

Now, we proceed by defining the finite-time Fourier transforms and
their inverses as follows:
\begin{align}
[\widetilde{U}(\omega_{n}), \widetilde{\eta}(\omega_{n})]&=\frac{1}{\tau} \int_{0}^{\tau} dt [U(t), \eta(t)] \exp(-i \omega_{n} t),  \\ 
\lbrack U(t), \eta(t)]&=\sum_{n=-\infty}^{\infty} [\widetilde{U}(\omega_{n}), \widetilde{\eta}(\omega_{n})] \exp(i \omega_{n} t),
\end{align}
with $\omega_{n}=2\pi n/\tau$.  In the frequency domain, the Gaussian
noise configurations denoted by $[{\eta(t):0<t<\tau}]$ is described by
the infinite sequence
$[\widetilde{\eta}(\omega_{n}):n=-\infty,...,-1,0,+1,...,\infty]$ of
Gaussian random variables with the correlation
\begin{equation}
\langle \widetilde{\eta}(\omega) \widetilde{\eta}^T(\omega^{\prime})\rangle
=\frac{2D}{\tau} \delta(\omega+\omega^{\prime})
\,\mathrm{diag} (1,\theta).
\end{equation}
In terms of the Fourier transform \eref{work3} can be written as
\begin{multline}
W_{\tau}=\frac{\tau}{2}\sum_{n=-\infty}^{\infty}
\biggl[
\frac{1}{ \tau_{0}^{2}}
\widetilde{U}^{T}(\omega_{n}){A}_{1}\widetilde{U}(-\omega_{n})\\
+ 
\frac{1}{ \tau_{0}}
\Bigl\{\widetilde{U}^{T}(\omega_{n}){A}_{2}^{T}\widetilde{\eta}(-\omega_{n})
+\widetilde{\eta}^{T}(\omega_{n}){A}_{2}\widetilde{U}(-\omega_{n})\Bigr\}
\biggr].
\end{multline}
\Eref{Langevin} gives
\begin{equation}
\widetilde{U}(\omega_{n}) =\tau_{0} {G} \widetilde{\eta} (\omega_{n}) - \frac{\tau_{0}}{\tau} {G} \Delta U,
\label{Utilde}
\end{equation}
 where ${G}=[iuI+{A}]^{-1}$ with $u=\omega_{n} \tau_{0}$, $\Delta
 U=U(\tau)-U(0)$, and $I$ being the identity matrix.  The elements of
 $G$ are: $G_{11}=(\delta+iu)^{-1}$, $G_{22}=(1+iu)^{-1}$,
 $G_{12}=\delta G_{11} G_{22}$, and $G_{21}=0$.  Note that
 ${G}(-u)={G}^*(u)$, $\widetilde{\eta}(-\omega)
 = \widetilde{\eta}^*(\omega)$, and $\widetilde{U}(-\omega)
 = \widetilde{U}^*(\omega)$. Substituting $\widetilde{U}$
 in \eref{work3}, and grouping the negative $n$ indices together with
 their positive counterparts in the summation, we get
\begin{align}
W_\tau=\Bigl[&
\frac{1}{2}\tau \widetilde{\eta}_{0}^{T}
\bigl(G_{0}^{T}A_{1}G_{0}+A_{2}G_{0}+G_{0}^{T}A_{2}^{T}\bigr) \widetilde{\eta}_{0}\notag\\
&-\Delta U^{T}\bigl(G_{0}^{T}A_{1}G_{0}
+G_{0}^{T}A_{2}^{T}\bigr)\widetilde{\eta}_{0} \notag\\
&+\frac{1}{2\tau}\Delta U^{T}\bigl(G_{0}^{T}A_{1}G_{0}\bigr) \Delta U\Bigr]\notag\\
+\sum_{n=1}^\infty\Bigl[&
\tau\widetilde{\eta}^{T}
\bigl(G^{T}A_{1}G^{*}+A_{2}G^{*}+G^{T}A_{2}^{T}\bigr) \widetilde{\eta}^{*} \notag
\\&- \widetilde{\eta}^{T}\bigl(G^{T}A_{1}G^{*}+A_{2}G^{*}\bigr)\Delta U \notag
\\&-\Delta U^{T}\bigl(G^{T}A_{1}G^{*}+G^{T}A_{2}^{T}\bigr)\widetilde{\eta}^{*} \notag \\
&+\frac{1}{\tau}~\Delta U^{T}\bigl(G^{T}A_{1}G^{*}\bigr) \Delta U
\Bigr],
\end{align}
in which ${G}_0={G}(u=0)={A}^{-1}$ and $\widetilde\eta_0=\widetilde\eta(0)$.

Next, we we express $U(\tau)$ in terms of the Fourier series
\begin{equation}
U(\tau)=\lim_{\epsilon \rightarrow0}\sum_{n=-\infty}^{\infty} \widetilde{U}(\omega_{n})
e^{-i\omega_{n}\epsilon}.
\end{equation}
While inserting $\widetilde{U}$ from \eref{Utilde} into the above
equation, we observe that $(1/\tau) \sum_{n} G
e^{-i\omega_{n}\epsilon}\rightarrow 0$ as
$\tau\rightarrow \infty$. This is because in the large-$\tau$ limit,
the summation can be converted to an integral which can be closed via
the lower half plane, and the $G$ is analytic there. Thus, using only
the first term of \eref{Utilde}, we get
\begin{align}
\sigma^T U(\tau) &= \tau_0 \sigma^T  G_0 \widetilde{\eta}_0\notag\\
&+\tau_0\sum_{n=1}^\infty 
\bigl[
e^{-i \omega_n \epsilon} \widetilde{\eta}^T G^T\sigma
+ e^{i \omega_n \epsilon}\sigma^T  G^* \widetilde{\eta}^*
\bigr].
\end{align}

Using this expression as well as $W_\tau$ from above in \eref{R-GF},
we get
\begin{equation}
Z(\lambda,U,\tau|U_{0})=\int \frac{d^{2}\sigma}{(2\pi)^{2}}
e^{i\sigma^{T}U} 
\prod_{n=0}^\infty\bigl\langle e^{s_n}\bigr\rangle, 
\label{R-GF2}
\end{equation}
where $s_n$ is  quadratic in $\widetilde\eta$, given by
\begin{equation}
s_{0}=-\frac{\lambda \tau}{2} \widetilde{\eta}_{0}^{T}B_{0}\widetilde{\eta}_{0}+\alpha_{0}^{T} \widetilde{\eta}_{0}-\frac{\lambda}{2\tau}\Delta
U^{T} G_{0}^{T}A_{1}G_{0} \Delta U,
\end{equation}
and
\begin{align}
s_{n}=&-\lambda \tau \widetilde{\eta}^{T}B_{n}\widetilde{\eta}^{*}+ \widetilde{\eta}^{T} \alpha_{n}+\alpha_{-n}^{T} \widetilde{\eta}^{*}\notag\\
&-\frac{\lambda}{\tau}\Delta
U^{T} G^{T}A_{1}G^{*} \Delta U
\quad\text{for}\quad n\ge 1,
\end{align}
in which we have used the following definitions:
\begin{align}
B_{n}&=G^{T}A_{1}G^{*}+A_{2}G^{*}+G^{T}A_{2}^{T}, \\
B_{0}&=G_{0}^{T}A_{1}G_{0}+A_{2}G_{0}+G_{0}^{T}A_{2}^{T},\\
\alpha_{n}&=\lambda \bigl[G^{T}A_{1}G^{*}+A_{2}G^{*} \bigr]\Delta U-i \tau_{0} e^{-i\omega_{n}\epsilon}G^{T}\sigma,\\
\alpha_{-n}^{T}&=\lambda \Delta
U^{T}\bigl[G^{T}A_{1}G^{*}+G^{T}A_{2}^{T}\bigr]-i \tau_{0}
e^{i\omega_{n}\epsilon} \sigma^{T}G^{*}.
\end{align}
Therefore, calculating the average $\langle e^{s_n}\rangle$
independently for each $n\ge 1$ with respect to the Gaussian PDF
$P(\widetilde\eta) =\pi^{-2}
(\det\Lambda)^{-1} \exp(-\widetilde{\eta}^{T}\Lambda^{-1}\widetilde{\eta}^{*})$
with $\Lambda=(2D/\tau) \mathrm{diag}(1,\theta)$
we get
\begin{equation}
\bigl\langle e^{s_{n}}\bigr\rangle
=\frac{\exp\bigl(\alpha_{-n}^{T}\Omega_{n}^{-1}\alpha_{n}
-\frac{\lambda}{\tau}\Delta U^{T}G^{T}A_{1}G^{*}\Delta U\bigr) 
}{\det(\Lambda\Omega_{n})},
\end{equation}
where $\Omega_{n}=\tau(\lambda  B_{n}+\tau^{-1}\Lambda^{-1})$. For the $n=0$
term, calculating the average $\langle e^{s_0}\rangle$ with respect to
the Gaussian PDF $P(\widetilde\eta_0) =(2\pi)^{-1}
(\det\Lambda)^{-1/2} \exp(-\frac{1}{2}\widetilde{\eta}_0^{T}\Lambda^{-1}\widetilde{\eta}_0)$
we get
\begin{equation}
\bigl\langle e^{s_{0}}\bigr\rangle
=\frac{\exp\bigl(\frac{1}{2}\alpha_{0}^{T}\Omega_{0}^{-1}\alpha_{0}
-\frac{\lambda}{2\tau}\Delta U^{T}G_0^{T}A_{1}G_0\Delta U\bigr) 
}{\sqrt{\det(\Lambda\Omega_{0})}}.
\end{equation} 

Since, $\langle e^{s_n}\rangle = \langle e^{s_{-n}}\rangle$, the
product in \eref{R-GF2} yields
\begin{multline}
\prod_{n=0}^\infty\bigl\langle e^{s_n}\bigr\rangle
=\exp\biggl(-\frac{1}{2}\sum_{n=-\infty}^{\infty} 
\ln\bigl[\det(\Lambda\Omega_{n})\bigr]\biggr)\\
\times 
\exp\biggl(\frac{1}{2\tau}\sum_{n=-\infty}^{\infty}
\bigl[\alpha_{-n}^{T}\tau\Omega_{n}^{-1}\alpha_{n}-\lambda\Delta
U^{T}G^{T}A_{1}G^{*}\Delta U\bigr]\biggr).
\label{prod<e^sn>}
\end{multline}
 The determinant in the above expression is found to be
\begin{equation}
 \det(\Lambda\Omega_{n})=[1+4\theta \lambda(1-\lambda)\delta^{2}u^{2}|G_{11}|^{2}|G_{22}|^{2}].  
\end{equation}
Now, taking the large $\tau$ limit, we replace the summations over $n$
 by integrals over $\omega$, i.e.,
 $\sum_{n}\rightarrow\tau\int \frac{d\omega}{2 \pi}$.  After,
 evaluating the integral, the argument of the exponential in first
 line of \eref{prod<e^sn>} yields
\begin{equation}
-\frac{\tau/ \tau_{0}}{4\pi} \int_{-\infty}^{\infty} du\, 
\ln\bigl[\det (\Lambda\Omega_{n})\bigr]=\tau\mu(\lambda),
\end{equation}
where $\mu(\lambda)$ is given by \eref{mu}. Similarly, converting the
argument of the exponential in the second line of \eref{prod<e^sn>} in
the integral forms, after some manipulation we get
\begin{equation}
\prod_{n=0}^\infty\bigl\langle e^{s_n}\bigr\rangle
\approx e^{\tau \mu(\lambda)} 
\exp\biggl[-\frac{1}{2} \sigma^{T}H_{1}\sigma+ i\Delta U^{T}H_{2}\sigma+ \frac{1}{2} \Delta U^{T}H_{3} \Delta U\biggr],
\label{prod<e^sn>2}
\end{equation}
in which $H_{1}$, $H_{2}$, and $H_{3}$ are given by
\begin{align}
H_{1}&=\frac{D\tau_{0}}{2 \pi} \int_{-\infty}^{\infty} du\,G^{*}
\widetilde\Omega^{-1} G^{T}, \\ 
H_{2}&=- \lim_{\epsilon\rightarrow
0} \frac{\lambda}{2\pi} \int_{-\infty}^{\infty}
du\,e^{iu\epsilon/\tau_0}\,
(G^{\dagger}\widetilde{A}_{1}G+G^{\dagger}\widetilde{A}_{2}^{T})
(\widetilde\Omega^{-1})^{*}G^{\dagger}, \\
H_{3}&=\frac{\lambda^{2}}{2\pi} \frac{1}{D\tau_{0}} 
\int_{-\infty}^{\infty}
du\,(G^{T}\widetilde{A}_{1}G^{*}+G^{T}\widetilde{A}_{2}^{T})
\widetilde\Omega^{-1}\notag\\
&\qquad\times(G^{T}\widetilde{A}_{1}G^{*}+\widetilde{A}_{2}G^{*}) 
-\frac{\lambda}{2 \pi}\frac{1}{D\tau_{0}} \int_{-\infty}^{\infty}
du~[G^{T}\widetilde{A}_{1}G^{*}],
\end{align}
where we have used where $\widetilde{\Omega}_n=\tau^{-1}D\Omega_n$ and
$\widetilde{A}_{1,2}=D A_{1,2}$ so that the integrands remain
dimensionless and dimensions are carried outside to the integrals. We
then evaluate the integrals performing the method of contours in the
complex $u$ plane, and using $G^{*}(u)+G(u)=2GAG^{*}$ and
$G^{*}(u)-G(u)=2 i u GG^{*}$, which yields
\begin{align}
H_{1}(\lambda) &= \frac{D\tau_{0}}{\delta(1+\delta)\nu(\lambda)} 
\begin{pmatrix}
1+\delta+\theta \delta^{2} & \theta \delta^{2}\\ \theta \delta^{2}
& \theta \delta+\theta \delta^{2}\\
\end{pmatrix},  \\
H_{2}(\lambda) &= - \frac{\nu(\lambda)-1}{2 \nu(\lambda)} \begin{pmatrix}
 1 & 0\\
 0 & 1\\  
\end{pmatrix}
- \frac{ \lambda \delta}{(1+\delta)\nu(\lambda)}
\begin{pmatrix}
 \theta\delta & \theta\delta\\
  1 & 0\\
\end{pmatrix},  \\
H_{3}(\lambda) &= \frac{\lambda \delta^{2}}{ D \tau_{0} (1+\delta) \nu(\lambda)} \begin{pmatrix}
\lambda \theta\delta & 1\\
 1 & \lambda-1\\
\end{pmatrix}.
\end{align}

Finally, inserting \eref{prod<e^sn>2} in \eref{R-GF2}, and performing
the Gaussian integral over $\sigma$ while using the facts that $H_1$
and $H_3$ are symmetric and $H_{3}= H_{1}^{-1}H_{2}^{T}+
H_{2}H_{1}^{-1}H_{2}^{T}$ we get
\begin{multline}
Z(\lambda,U,\tau|U_{0}) \approx e^{\tau \mu(\lambda)} 
\exp\left(-\frac{1}{2} U_{0}^{T}L_{2}(\lambda)U_{0}\right)\\
\times\frac{1}{2\pi \sqrt{\det
 H_{1}(\lambda)}} \exp\left(-\frac{1}{2}U^{T}L_{1}(\lambda)U\right),
\label{R-GF3}
\end{multline}
with $L_{1}(\lambda) = H_{1}^{-1}+H_{1}^{-1}H_{2}^{T}$ and
$L_{2}(\lambda) = -H_{1}^{-1} H_{2}^{T}$. From the above equation, it
is trivial to identify $\chi(U_0,\lambda)$ and $\Psi(U,\lambda)$ used
in \eref{characteristic.1}. Since, $L_1+L_2=H_1^{-1}$, it is evident
that $\int \chi(U,\lambda)\Psi(U,\lambda)\, dU=1$.

Application of the Langevin operator given by \eref{langevin operator}
on $\Psi(U,\lambda)$ yields
\begin{align}
\mathcal{L_{\lambda}} \psi(U,\lambda)
&=\biggl[D\bigl(L_{1}^{1,1}\bigr)^{2}+\alpha D\bigl(L_{1}^{1,2}\bigr)^{2}-
\frac{\delta}{\tau_{0}}L_{1}^{1,1}\biggr]x^{2}\psi(U,\lambda)\notag\\
&+ \biggl[D\bigl(L_{1}^{1,2}\bigr)^{2}+\alpha
D\bigl(L_{1}^{2,2}\bigr)^{2}+\frac{\delta}{\tau_{0}}(1-2\lambda)L_{1}^{1,2} 
\notag\\
&\qquad-\frac{1}{\tau_{0}}L_{1}^{2,2}
-\frac{\delta^{2}}{D\tau_{0}^{2}}\lambda(1-\lambda)\biggr]
y^{2}\psi(U,\lambda)
\notag\\
&+\biggl[2D
L_{1}^{1,1}L_{1}^{1,2}
+ 2D \alpha L_{1}^{1,2}L_{1}^{2,2}
+\frac{\delta}{\tau_{0}}(1-2\lambda)L_{1}^{1,1}\notag\\
&\qquad-\frac{1+\delta}{\tau_{0}}L_{1}^{1,2}
+\frac{\delta^{2}}{D\tau_{0}^{2}}\lambda\biggr]
xy\psi(U,\lambda)\notag\\
&+\biggl[-D L_{1}^{1,1}-\alpha
DL_{1}^{2,2}+\frac{1+\delta}{\tau_{0}}\biggr]\psi(U,\lambda),
\end{align}
where $L_1^{i,j}$ denotes the $(i,j)$-th element of the matrix
$L_1$. Using the explicit expressions on the right-hand side of the
above equation, after simplification, we find the coefficients of
$x^2 \Psi(U,\lambda)$, $y^2 \Psi(U,\lambda)$, and $xy \Psi(U,\lambda)$
to be zero. The last term in square brackets in front of
$\Psi(U,\lambda)$ yields $\mu(\lambda)$ given by \eref{mu}. This
verifies the eigenvalue equation $\mathcal{L}_\lambda \Psi (U,\lambda)
= \mu(\lambda) \Psi(U,\lambda)$.

The steady-state of the system is given by 
\begin{equation}
P_\text{SS}
 (U)=\Psi(U,0)=\frac{\exp\left(-\frac{1}{2}U^{T}H_{1}^{-1}(0)U\right)}
{2\pi \sqrt{\det  H_{1}(0)}}.
\end{equation}
Integrating \eref{R-GF} over $U$ and then averaging over the initial
condition $U_{0}$ with respect to the steady state distribution
$P_\text{SS}(U_{0})$, we obtain $Z(\lambda)$ given
by \eref{Z-asymptotic}, with
\begin{align}
g(\lambda)&=\bigl(\det H_{1}(\lambda) \det H_{1}(0) \det L_{1}(\lambda) 
\det [H_{1}^{-1}(0)+L_{2}(\lambda)]\bigr)^{-1/2} \notag\\
&=\bigl(\det [1- \nu(\lambda) H_{2}^{T}(\lambda)]
\det[1+ H_{2}^{T}(\lambda)]\bigr) ^{-1/2},
\end{align}
where to obtain the second expression, we have substituted the
expressions of $L_1$, $L_2$ and $H_1(0)=\nu(\lambda) H_1(\lambda)$.
Inserting the matrix $H_2$ and evaluating the determinants, after
simplification, we obtain \eref{g}.

\section{Singularities of $g(\lambda)$}
\label{singularities of g}

\begin{figure}
\includegraphics[width=\hsize]{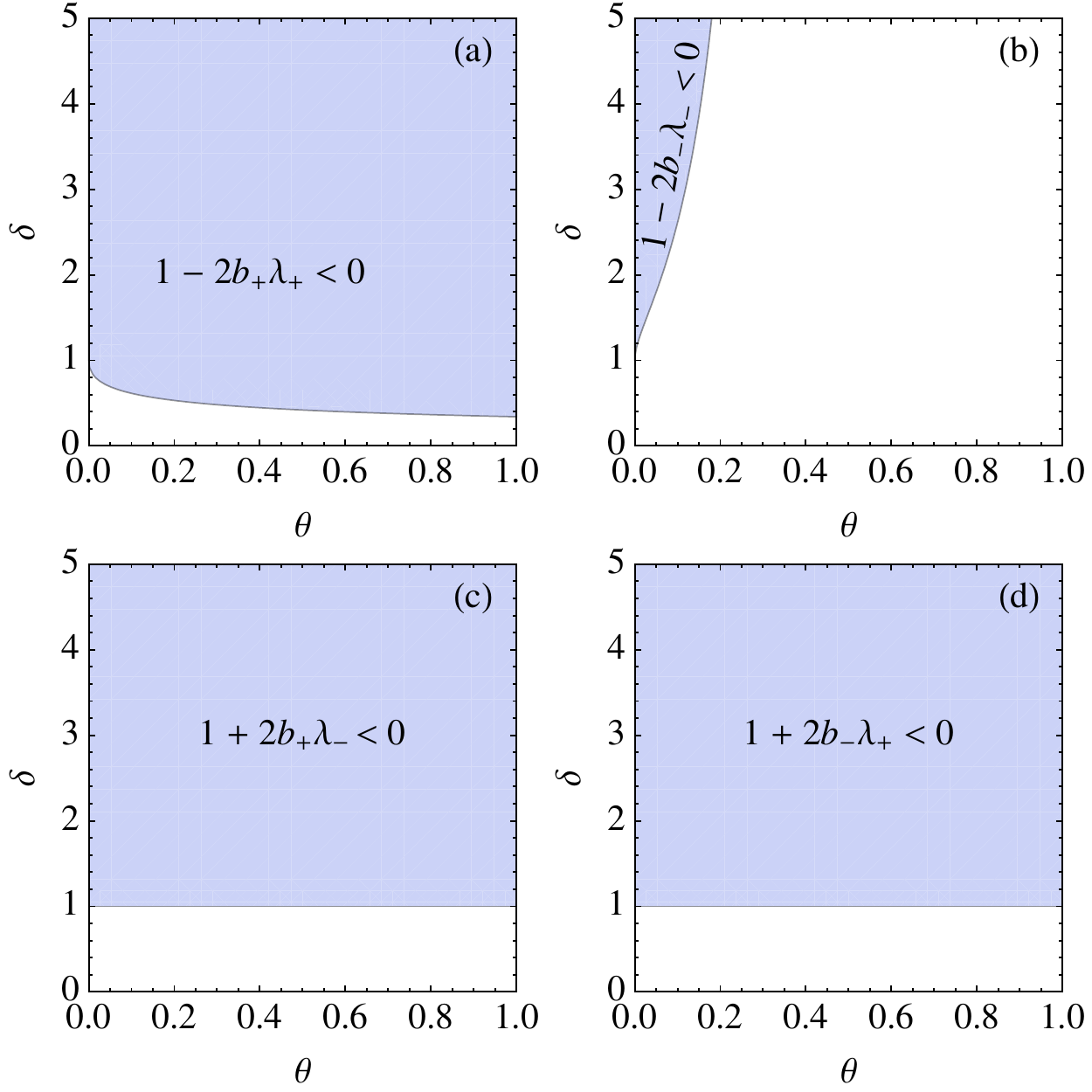}
\caption{\label{phase}(Color online) In the
shaded regions of the $(\theta,\delta)$ plane in the figures (a), (b),
(c) and (d), the respective mathematical conditions given there are
satisfied and consequently $g(\lambda)$ possesses singularities at
$\lambda_\text{a}$, $\lambda_\text{b}$, $\lambda_\text{c}$, and
$\lambda_\text{d}$ respectively, given by
Eqs.~\eqref{lambda-a}--\eqref{lambda-d}.}
\end{figure}

From \eref{nu-2} we recall that $\nu(\lambda_\pm)=0$ and
$\nu(\lambda)>0$ (is a semicircle) for
$\lambda\in(\lambda_-,\lambda_+)$. Moreover, all the four functions
$1\pm 2b_{\pm}\lambda$ are linear in $\lambda$ with slopes $\pm
2b_\pm$ (where all four combinations of the two $\pm$ signs are
considered).  Therefore, for example, if $(1-2b_{+}\lambda)$ has
opposite signs at the two end points $\lambda_\pm$, then the function
$[\nu(\lambda)+(1 -2 b_{+}\lambda)]$ must cross zero at some
intermediate $\lambda$. This is also true for the other three
cases. From Eqs.~\eqref{b_pm} and \eqref{lambda_pm} respectively, we
note that $b_+>0$, $b_-<0$ and $\lambda_+>0$, $\lambda_-<0$.  One can
therefore determine whether $g(\lambda)$ has a singularity as follows
(see \fref{phase}):
\begin{enumerate}[(a)]
\item  Evidently,  $1-2b_{+}\lambda_- >0$. Thus, 
$\nu(\lambda_\text{a})+ 1 -2b_{+}\lambda_\text{a} =0$ for a specific
$\lambda_\text{a}\in(\lambda_-,\lambda_+)$ if and only if
$1-2b_{+}\lambda_+ <0$. When this happens [see \fref{phase}~(a)], the
position of the singularity can be found as
\begin{equation}
\lambda_\text{a}= (a + b_+)/(a+b_+^2).
\label{lambda-a}
\end{equation}
It is evident that $\lambda_\text{a} >0$.

\item  Evidently, $1-2b_{-}\lambda_+ >0$. Thus, $\nu(\lambda_\text{b})+ 1
-2b_{-}\lambda_\text{b} =0$ for a specific
$\lambda_\text{b}\in(\lambda_-,\lambda_+)$ if and only if
$1-2b_{-}\lambda_- <0$.  When this happens [see \fref{phase}~(b)], the
position of the singularity can be found as
\begin{equation}
\lambda_\text{b}=(a + b_-)/(a+b_-^2)
\label{lambda-b}
\end{equation}
and it can be shown that $\lambda_\text{b} <0$.

\item   Evidently, $1+2b_{+}\lambda_+ >0$. Thus, $\nu(\lambda_\text{c})+ 1
+2b_{+}\lambda_\text{c} =0$ for a specific $\lambda_\text{c}\in
(\lambda_-,\lambda_+)$ if and only if $1+2b_{+}\lambda_- <0$.  When
this happens [see \fref{phase}~(c)], the position of the singularity
can be found as
\begin{equation}
\lambda_\text{c}=(a - b_+)/(a+b_+^2)
\label{lambda-c}
\end{equation}
and it can be shown that $\lambda_\text{c} < 0$.

\item   Evidently, $1+2b_{-}\lambda_- >0$. Thus, $\nu(\lambda_\text{d})+ 1 +
2b_{-}\lambda_\text{d} =0$ for a specific
$\lambda_\text{d}\in(\lambda_-,\lambda_+)$ if and only if
$1+2b_{-}\lambda_+ <0$.  When this happens [see \fref{phase}~(d)], the
position of the singularity can be found as
\begin{equation}
\lambda_\text{d}=(a - b_-)/(a+b_-^2).
\label{lambda-d}
\end{equation}
It is evident that $\lambda_\text{d} >0$. Moreover, it can be shown
that $\lambda_\text{c}+\lambda_\text{d}=1$.
\end{enumerate}
It is easily seen that the singularities of $g(\lambda)$ are branch
points (square root singularities) and the function $f_w(\lambda)$ at
these singularities is given by
\begin{equation}
h_i(w):=f_w(\lambda_i)= \frac{1}{2}\bigl[1-\nu(\lambda_i) \bigr]
+\lambda_i w,
\label{f_i}
\end{equation}
where the index $i$ stands for one of the indices from the set
$\{\text{a, b, c, d}\}$. Substituting $\nu(\lambda_i)$ at the
singularities using the conditions from above, we get
\begin{align}
\label{ha(w)}
h_\text{a}(w)&=(1-b_+\lambda_\text{a})+\lambda_\text{a} 
w,\\
\label{hb(w)}
h_\text{b}(w)&=(1-b_-\lambda_\text{b})+\lambda_\text{b} 
w,\\
\label{hc(w)}
h_\text{c}(w)&= (1+b_+\lambda_\text{c})+\lambda_\text{c} 
w,\\
\label{hd(w)}
h_\text{d}(w)&= (1+b_-\lambda_\text{d})+\lambda_\text{d} w.
\end{align}
It is also useful to define the non-singular part of $g(\lambda)$ at a
singularity as
\begin{equation}
\widetilde{g}(\lambda_i)=\lim_{\lambda\rightarrow\lambda_i}
\bigl|(\lambda-\lambda_i)^{1/2}  g(\lambda)\bigr|.
\label{g-ns}
\end{equation}

We note that the for a given set of parameters $\theta$ (or $\alpha$)
and $\delta$, the position of the singularities (whenever they exist)
are fixed within the interval $(\lambda_-,\lambda_+)$.  The specific
values of $w$ at which the saddle point coincides with one of the
singularities is obtained by solving $\lambda^*(w^*_i)=\lambda_i$ as
\begin{equation}
w^*_i = \frac{(1-2\lambda_i)\sqrt{a}}{\sqrt{(1+1/a) -
(2\lambda_i-1)^2}}.
\label{wstar}
\end{equation}
Since, $(1+1/a)=(2\lambda_\pm-1)^2$ and $\lambda_-< \lambda_i
<\lambda_+$, the term under the square root in the above equation is
always positive.

\subsection{The case: $\delta<1$}

For any $\delta<1$, there exists a $\theta_{c_1}$ given by the
solution of $1-2b_+\lambda_+=0$ as
\begin{equation}
\theta_{c_1}(\delta)=
\frac{\left(1-\delta ^2\right)^2}{\delta ^2 \left(3+10 \delta
+3 \delta ^2\right)}, 
\label{theta_c1}
\end{equation}
 and for $\theta<\theta_{c_1}$ the
function $g(\lambda)$ has no singularities whereas it has one
singularity for $\theta>\theta_{c_1}$. As $\delta\rightarrow 0$ we get
$\theta_{c_1} \simeq 1/(3 \delta^2)$ whereas $\theta_{c_1} \simeq
(1-\delta)^2/4$ as $\delta\rightarrow 1^-$.

The $\theta=\theta_{c_1}(\delta)$ line corresponds to the
$\alpha=\alpha_{c_1}(\delta)$ line in the $(\alpha,\delta)$ plane,
where
\begin{equation}
\alpha_{c_1}(\delta)=
\frac{(1+\delta)\left(1-\delta \right)^2}{3+10 \delta
+3 \delta ^2}.
\label{alpha_c1}
\end{equation}

\subsection{The case: $\delta>1$}

For $\delta>1$, there again exists a $\theta_{c_2}$ given by the
solution of $1-2b_-\lambda_-=0$ as
\begin{equation}
\theta_{c_2}(\delta)=\frac{\left(\delta ^2-1\right)^2}{\delta
^2 \left(3+10 \delta +3 \delta ^2\right)},
\label{theta_c2}
\end{equation}
and $g(\lambda)$ has either three or four singularities depending on
whether $\theta >\theta_{c_2}$ or $\theta <\theta_{c_2}$. In the limit
$\delta\rightarrow\infty$ we get $\theta_{c_2}=1/3$ and
$\theta_{c_2}\rightarrow 0$ as $\delta\rightarrow 1$. More precisely,
$\theta_{c_2}\simeq 1/3-10/(9\delta)$ as $\delta\rightarrow\infty$,
whereas $\theta_{c_2}\simeq (\delta-1)^2/4$ as $\delta\rightarrow
1^+$.

The $\theta=\theta_{c_2}(\delta)$ line corresponds to the
$\alpha=\alpha_{c_2}(\delta)$ line in the $(\alpha,\delta)$ plane,
where
\begin{equation}
\alpha_{c_2}(\delta)=
\frac{(1+\delta)\left(\delta -1 \right)^2}{3+10 \delta
+3 \delta ^2}.
\label{alpha_c2}
\end{equation}

\subsection{The case: $\delta=1$}

It is instructive to illustrate the particular case of $\delta=1$, for
which we have $\alpha=\theta/2$ and $a=\theta/4$. Here from
Eqs.~\eqref{b_pm} and \eqref{lambda_pm} we get
$2b_\pm=\theta\lambda_\pm$ and $\theta\lambda_+\lambda_-=-1$.  It
follows that:
\begin{enumerate}[(a)]
\item $1-2b_+\lambda_-=2$ and $1-2b_+\lambda_+=-\theta\lambda_+ <0$ for
$\theta>0$. This implies $g(\lambda)$ has a singularity at
$\lambda=\lambda_\text{a}$. We get
\begin{math}
\lambda_\text{a}=(1+2\lambda_+)/(2+\theta\lambda_+)
\end{math}
and $\lambda_\text{a}\in (0,\lambda_+)$.

\item $1-2b_-\lambda_+=2$ and $1-2b_-\lambda_-=-\theta\lambda_- >0$ for
$\theta>0$. This implies $g(\lambda)$ does not have any singularity at
$\lambda=\lambda_\text{b}$. 

\item $1+2b_+\lambda_+=2+\theta\lambda_+ >0$ and
$1+2b_+\lambda_-=0$. However, since $\nu(\lambda_-)=0$, $g(\lambda)$
has a singularity at $\lambda=\lambda_\text{c}=\lambda_-$.

\item $1+2b_-\lambda_+=2+\theta\lambda_- >0$ as
$\theta\lambda_-\in(-1,0)$. Moreover, $1+2b_-\lambda_+=0$ and
$\nu(\lambda_+)=0$. Therefore, $g(\lambda)$ has a singularity at
$\lambda=\lambda_\text{d}=\lambda_+$.
\end{enumerate}
However, we have already seen that $\lambda^*\rightarrow \lambda_\pm$
only when $w\rightarrow\mp\infty$. Therefore, for all practical
purposes (any finite $w$) the singularities at $\lambda_\pm$ are not
relevant and hence we treat this case together with the case $\delta
<1$, $\theta>\theta_{c_1}$ where $g(\lambda)$ has only one
singularity. However, for the $\delta=1$ case, in principle, one can
also use the results of \sref{three-singularities}, where the case of
the three singularities is discussed.

\section{Steepest descent method with a branch point}
\label{steepest-descent}

Let us consider the integral
\begin{equation}
I=\frac{1}{2\pi i}\int_{-i\infty}^{i\infty}g_1(\lambda)\frac{e^{\tau f_w(\lambda)}}{\sqrt{\lambda_\text{a}-\lambda}}
\, d\lambda,
\label{integral}
\end{equation}
where $\lambda_\text{a} >0$.  The position of the saddle point
$\lambda^*$ depends on the value of $w$, and depending on whether
$w>w^*_\text{a}$ or $w< w^*_\text{a}$ we have $\lambda^*
< \lambda_\text{a}$ or $\lambda^* > \lambda_\text{a}$ respectively.
In the following, we consider the two cases one by one.

\subsection{The branch point is not between the origin and the saddle point:
$\lambda_\text{a}\not\in(0,\lambda^*)$}
\label{sec: branch point outside}

\begin{figure}
\includegraphics[width=.85\hsize]{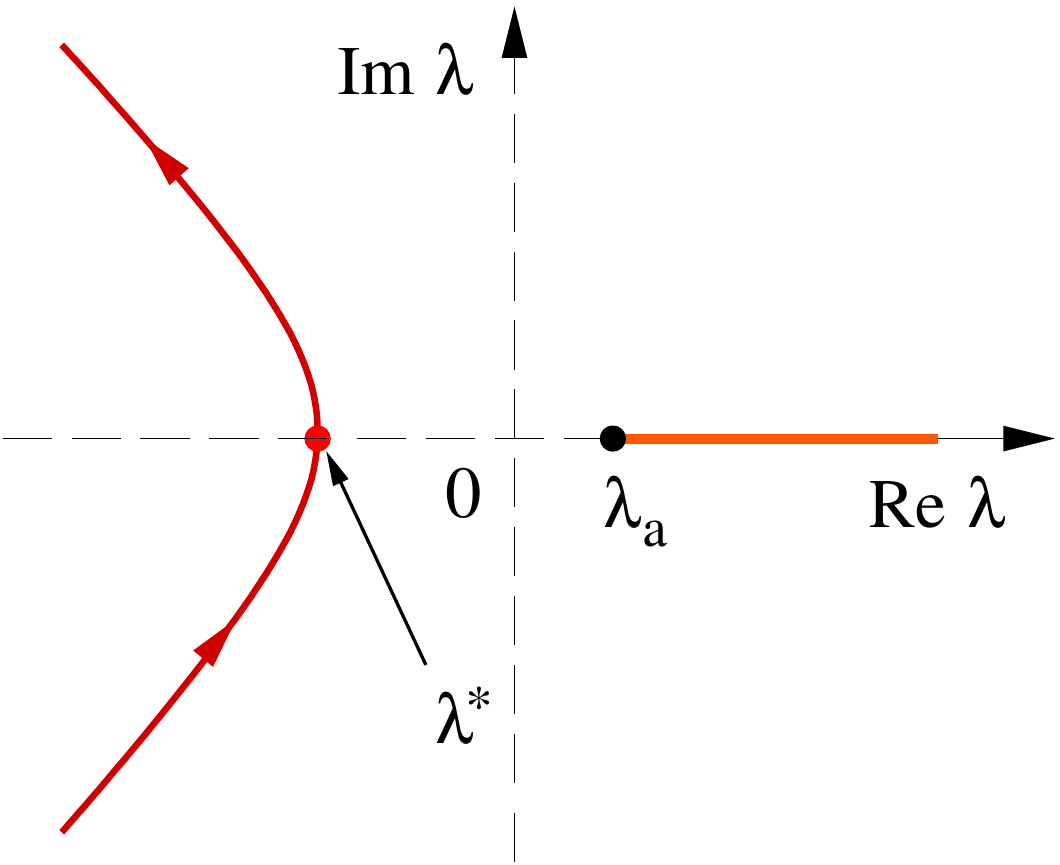}
\caption{\label{steepest-descent contour1a}  (Color online)  
Schematic steepest descent contour (in red) for the case when the
branch point $\lambda_\text{a}$ is not between the origin and the
saddle point $\lambda^*$. Here, it is shown for $\lambda^* <0$,
however, one can also have $\lambda^*\in (0,\lambda_\text{a})$. The
direction towards which the contour bends, depends on the value of
$w$. Here, it is shown for $w <0$. For $w>0$ the contour bends towards
the positive $\mathrm{Re}(\lambda)$ axis, whereas for $w=0$, the
steepest descent contour is parallel to the $\mathrm{Im}(\lambda)$
axis. The thick solid (orange) line along the $\mathrm{Re}(\lambda)$
axis from $\lambda_\text{a}$ represents the branch cut.}
\end{figure}

In this case, since $\lambda_\text{a}$ lies outside the interval
$(0,\lambda^*)$, one can deform the contour of integration
in \eref{integral} into the steepest descent path through $\lambda^*$
without hitting $\lambda_\text{a}$ (see \fref{steepest-descent
contour1a}). Along the steepest descent contour we define
\begin{equation}
f_w(\lambda) - f_w(\lambda^*) = -u^2.
\label{u-saddle}
\end{equation}
Therefore, $\lambda_\text{a}$ is mapped to a branch point at $u=-ib$
with
\begin{equation}
b=\sqrt{f_w(\lambda_\text{a}) - f_w(\lambda^*)} 
\label{b}
\end{equation}
and \eref{integral} becomes
\begin{equation}
I=\frac{e^{\tau f_w(\lambda^*)}}{2\pi i}\int_{-\infty}^\infty
q_1(u)\, \frac{e^{-\tau u^2}}{\sqrt{b-iu}}\, du
\end{equation}
with 
\begin{equation}
q_1(u)=g_1(\lambda)\,\frac{\sqrt{b-iu}}{\sqrt{\lambda_\text{a}-\lambda}} 
\,\frac{d\lambda}{du}.
\end{equation}
Now, making a change of variable $\sqrt{\tau} u\rightarrow u$ and
taking the large-$\tau$ limit we get 
\begin{equation}
I\approx\frac{e^{\tau f_w(\lambda^*)}}{2\pi i} q_1(0) \tau^{-1/4} \int_{-\infty}^\infty
 \frac{e^{-u^2}}{\sqrt{b\sqrt{\tau}-iu}}\, du,
\end{equation}
where 
\begin{equation}
q_1(0)=g_1(\lambda^*)\,
\frac{\sqrt{b}}{\sqrt{\lambda_\text{a}-\lambda^*}} 
\,\frac{d\lambda}{du}\bigg|_{\lambda\rightarrow\lambda^*}.
\end{equation}
Using 
\begin{math}
-u^2=\frac{1}{2} f''(\lambda^*)
(\lambda-\lambda^*)^2 +\dotsb
\end{math}
as $\lambda\rightarrow\lambda^*$, it can be found that
\begin{equation}
\frac{d\lambda}{du}\bigg|_{\lambda\rightarrow\lambda^*}
=\frac{i\sqrt2}{\sqrt{f''(\lambda^*)}} \, .
\label{jacobian}
\end{equation}
Therefore, we get
\begin{equation}
I\approx
\frac{g_1(\lambda^*)}
{\sqrt{\lambda_\text{a}-\lambda^*}} 
\frac{e^{\tau f_w(\lambda^*)}}
{\sqrt{2\pi\tau f''_w(\lambda^*)}}\,
R_1(\sqrt\tau\, b),
\end{equation}
where 
\begin{equation}
R_1(z) = \sqrt\frac{z}{\pi} \int_{-\infty}^\infty
  \frac{e^{-u^2}}{\sqrt{z-iu}}\, du.
\label{R_1 integral}
\end{equation}
We perform this integral in the Mathematica to get \eref{R_1}.

\subsection{The branch point is between the origin and the saddle point:
$\lambda_\text{a}\in(0,\lambda^*)$}
\label{sec: branch point inside}

\begin{figure}
\includegraphics[width=\hsize]{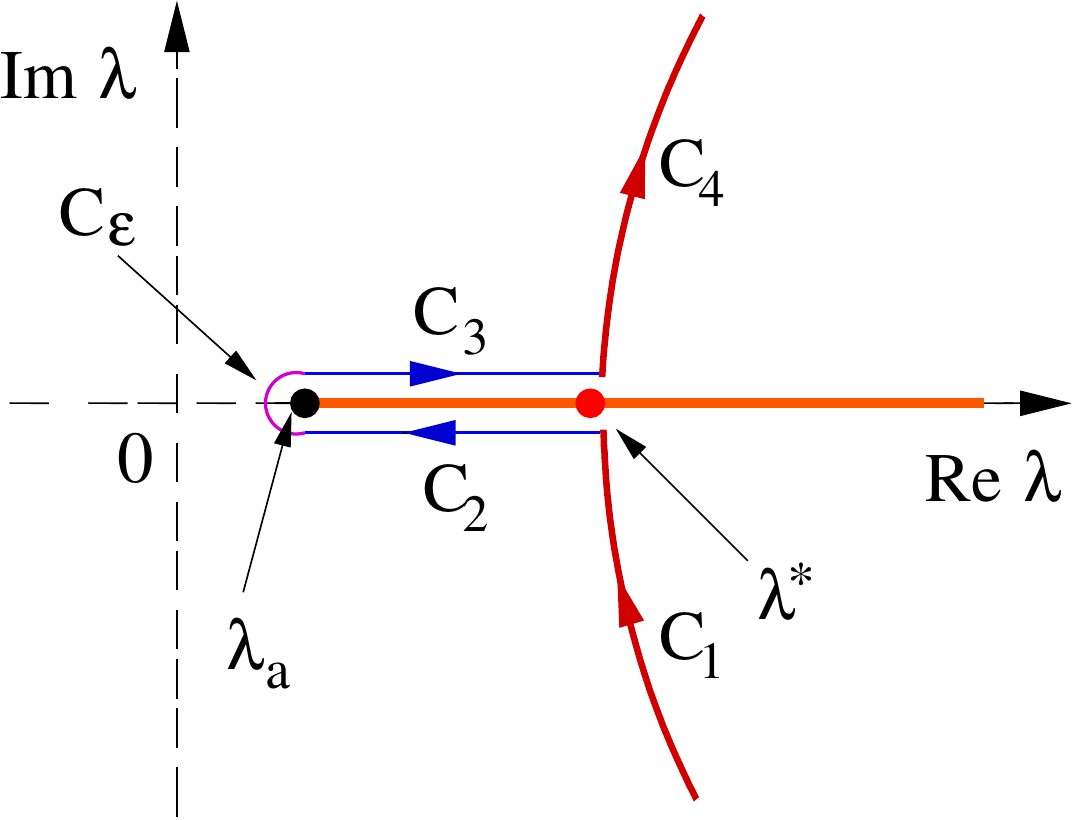}
\caption{\label{steepest-descent contour1b} (Color online) 
Schematic steepest descent contour for the case when the branch point
$\lambda_\text{a}$ is between the origin and the saddle point
$\lambda^*$.  The thick solid (orange) line along the
$\mathrm{Re}(\lambda)$ axis from $\lambda_\text{a}$ represents the
branch cut. The steepest descent contour goes around the branch cut as
shown by $C_2$ and $C_3$ (in blue).  The contribution coming from the
circular contour (in magenta) $C_\epsilon$ around the branch point
becomes zero in the the limit of the radius $\epsilon\rightarrow 0$.
The direction towards which the contours $C_1$ and $C_4$ (shown in
red) bend, depends on the value of $w$. Here, it is shown for $w >
0$. For $w<0$ the $C_1$ and $C_4$ bend towards the negative
$\mathrm{Re}(\lambda)$ axis, whereas for $w=0$, they are parallel to
the $\mathrm{Im}(\lambda)$ axis. }
\end{figure}

In this case, since $\lambda_\text{a}$ lies in the interval
$(0,\lambda^*)$, the deformed contour the through $\lambda^*$ wraps
around the branch cut. The contour of integration $C=C_1+C_2+C_3+C_4
+C_\epsilon$ is shown in \fref{steepest-descent contour1b}. The contour
$C_\epsilon$ represents the circular contour of radius $\epsilon$
going around the branch point, and its contribution becomes zero in
the limit $\epsilon\rightarrow 0$.  The integral in \eref{integral}
can be written as
\begin{math}
I=P_\text{B} + P_\text{S},
\end{math}
where $P_\text{B} (w,\tau)$ is the contribution coming from the
integrations along the contours $C_2$ and $C_3$, whereas $P_\text{S}$
is the saddle point contribution coming from the integrations along
the contours $C_1$ and $C_4$. In the following we evaluate
$P_\text{B}$ and $P_\text{S}$.

\subsubsection{Branch cut contribution}
\label{sec: Branch cut contribution}

We consider,
\begin{equation}
P_\text{B}=\frac{1}{2\pi i}\int_{C_2+C_3}
g_1(\lambda)\frac{e^{\tau
f_w(\lambda)}}{\sqrt{\lambda_\text{a}-\lambda}}
\, d\lambda.
\label{integral-branch}
\end{equation}
We note that $\sqrt{\lambda_\text{a}-\lambda}$ changes when one goes
from $C_2$ to $C_3$. More precisely,
$\lambda_\text{a}-\lambda=|\lambda_\text{a}-\lambda| e^{i\phi}$, where
$\phi=+\pi$ on $C_2$ and $\phi=-\pi$ on $C_3$ (as $\phi=0$ for
$\lambda <\lambda_\text{a}$ on the real-$\lambda$ line). Therefore,
$\sqrt{\lambda_\text{a}-\lambda}=+i|\lambda_\text{a}-\lambda|^{1/2}$
on $C_2$ and
$\sqrt{\lambda_\text{a}-\lambda}=-i|\lambda_\text{a}-\lambda|^{1/2}$
on $C_3$, using which from \eref{integral-branch} we get
\begin{equation}
P_\text{B}=\frac{1}{\pi}\int_{\lambda_\text{a}}^{\lambda^*}
g_1(\lambda)\frac{e^{\tau
f_w(\lambda)}}{|\lambda-\lambda_\text{a}|^{1/2}}
\, d\lambda.
\label{integral-branch2}
\end{equation}

Since $f_w(\lambda)$ is real, $f_w(\lambda_\text{a}) > f_w(\lambda) >
f_w(\lambda^*)$ for $\lambda_\text{a} < \lambda < \lambda^*$, and
$f_w(\lambda)$ is minimum at $\lambda^*$ along the real $\lambda$
line, we set
\begin{equation}
f_w(\lambda)- f_w(\lambda_\text{a}) = -2bu + u^2.
\label{mapping branch}
\end{equation}
The branch point $\lambda_\text{a}$ is mapped to $u=0$.  Using
$f'_w(\lambda^*)=0$, we find that the saddle point is mapped to $u=b$,
and $b$ can be found by putting $\lambda=\lambda^*$ and $u=b$ in the
above equation, which gives \eref{b}.  With the above mapping from
$\lambda$ to $u$, \eref{integral-branch} becomes
\begin{equation}
P_\text{B}=\frac{e^{\tau f_w(\lambda_\text{a})}}{\pi}
\int_0^{b} q_2(u) \frac{e^{-\tau (2bu-u^2)}}{\sqrt{u}}\, du,
\label{integral-branch3}
\end{equation}
where
\begin{equation}
q_2(u)=g_1(\lambda)\,\frac{\sqrt{u}}{|\lambda-\lambda_\text{a}|^{1/2}} 
\,\frac{d\lambda}{du}.
\end{equation}
From \eref{mapping branch}, we get
\begin{equation}
\frac{d\lambda}{du} = \frac{2(u-b)}{f'_w(\lambda)},
\label{jacobian2}
\end{equation}
which is finite and nonzero everywhere between $u=0$ and $u=b$. Near
$u=0$ we get
\begin{equation}
\left.\frac{d\lambda}{du}\right|_{u=0} = \frac{2b}{-f'_w(\lambda_\text{a})}.
\end{equation}
On the other hand, near $u=b$, by applying  L'Hospital rule
to \eref{jacobian2} we get
\begin{equation}
\left.\frac{d\lambda}{du}\right|_{u=b} = \frac{\sqrt2}{\sqrt{f''_w(\lambda^*)}}.
\end{equation}

Now, making a change of variable $\sqrt{\tau} u\rightarrow u$
in \eref{integral-branch3} and then taking the large-$\tau$ limit we
get
\begin{equation}
P_\text{B}\approx \frac{e^{\tau f_w(\lambda_\text{a})}}{\pi}
 \frac{q_2(0)}{\tau^{1/4}}  R_2(\sqrt\tau b),
\label{integral-branch4}
\end{equation}
where 
\begin{equation}
\label{R2}
R_2(z)=\int_0^z \frac{1}{\sqrt{u}}\, e^{-2zu+u^2}\, du.
\end{equation}
The asymptotic forms of $R_2(z)$ can be easily determined from the
above integral, which gives $R_2(z)\sim \sqrt{\pi}/\sqrt{2 z}$ as
$z\rightarrow\infty$.

It can be shown that
\begin{equation}
\frac{\sqrt{u}}{|\lambda-\lambda_\text{a}|^{1/2}} 
\,\frac{d\lambda}{du} 
\xrightarrow[\lambda\rightarrow\lambda_\text{a}]{u\rightarrow 0}
\Biggl[\frac{d\lambda}{du}\bigg|_{u=0}\Biggr]^{1/2}.
\end{equation}
Therefore,
\begin{equation}
q_2(0)=g_1(\lambda_\text{a})\,
\Biggl[\frac{d\lambda}{du}\bigg|_{u=0}\Biggr]^{1/2}.
\end{equation}

\subsubsection{Saddle point contribution}
\label{sec: Saddle point contribution}

We consider,
\begin{equation}
P_\text{S}=\frac{1}{2\pi i}\int_{C_1+C_4} g_1(\lambda)\frac{e^{\tau f_w(\lambda)}}{\sqrt{\lambda_\text{a}-\lambda}}
\, d\lambda.
\label{integral-saddle}
\end{equation}
We make a transform from $\lambda$ to $u$ as defined
by \eref{u-saddle}. In this case, the branch point $\lambda_\text{a}$
is mapped to a branch point at $u=ib$ where $b$ is given by \eref{b},
and \eref{integral-saddle} becomes
\begin{equation}
P_\text{S}=\frac{e^{\tau f_w(\lambda^*)}}{2\pi i}\int_{-\infty}^\infty
q_3(u)\, \frac{e^{-\tau u^2}}{\sqrt{b+iu}}\, du
\end{equation}
with 
\begin{equation}
q_3(u)=g_1(\lambda)\,\frac{\sqrt{b+iu}}{\sqrt{\lambda_\text{a}-\lambda}} 
\,\frac{d\lambda}{du}.
\end{equation}
We found in the preceding sub-subsection that
$\sqrt{\lambda_\text{a}-\lambda^*}=\pm
i|\lambda_\text{a}-\lambda^*|^{1/2}$ below ($+$) and above ($-$) the
branch cut respectively. Therefore, $q_3(u)$ approaches two different
limits as $u\rightarrow 0$ form above ($0^+$) and below ($0^-$)
respectively:
\begin{equation}
q_3(0^\pm)=\mp \frac{g_1(\lambda^*)\sqrt{b}}
{|\lambda_\text{a}-\lambda^*|^{1/2}} 
\frac{\sqrt2}{\sqrt{f''(\lambda^*)}} \, ,
\end{equation}
where we have used 
\eref{jacobian} for the Jacobian. Thus, upon changing $\sqrt\tau
u\rightarrow u$ and taking the large-$\tau$ limit yields
\begin{equation}
P_\text{S}\approx
\frac{g_1(\lambda^*)}
{|\lambda_\text{a}-\lambda^*|^{1/2}} 
\frac{e^{\tau f_w(\lambda^*)}}{\sqrt{2\pi\tau f''(\lambda^*)}}\,
R_4(\sqrt{\tau}\, b),
\label{P_S}
\end{equation}
where
\begin{align}
R_4(z) &= \sqrt\frac{z}{\pi}\left[ \int_{0}^\infty
  \frac{e^{-u^2}\, du}{\sqrt{z+iu}} - \int_{-\infty}^0
  \frac{e^{-u^2} \, du}{\sqrt{z+iu}}
\right]i\notag\\
&= \sqrt\frac{z}{\pi}
\int_{0}^\infty du\,e^{-u^2}
\left[
  \frac{1}{\sqrt{z+iu}} - 
  \frac{1}{\sqrt{z-iu}}
\right]i .
\label{R_4 integral}
\end{align}
We evaluate this integral in Mathematica to get \eref{R4}, where the
generalized hypergeometric function has the series expansion
\begin{equation}
{}_2F_2 (a_1,a_2; b_1,b_2; z)=\sum_{n=0}^{\infty } \frac{\left(a_1\right)_n
\left(a_2\right)_k}{\left(b_1\right)_k \left(b_2\right)_n} \frac{z^n}{n!}
\label{2F2}
\end{equation}
with $(a)_n=a (a+1)(a+2)\cdots(a+n-1)$, $(a)_0=1$ being the the
Pochhammer symbol.

The large $z$ behavior of $R_4(z)$ can be found by expanding the term
inside the square bracket in \eref{R_4 integral} in powers of
$1/z$ and integrating term by term. This gives
\begin{math}
R_4(z) \simeq 1/(2\sqrt{\pi}\, z) 
\end{math}
for large $z$. 

On the other hand, $R_4(z) \simeq \Gamma(1/4) \sqrt{z/2\pi}$ for small
$z$.  Using this together with
$\lim_{\lambda^*\rightarrow\lambda_\text{a}}\sqrt{b}/
|\lambda_\text{a}-\lambda^*|^{1/2}= [f''(\lambda^*)/2]^{1/4}$
in \eref{P_S} we get
\begin{equation}
P_\text{S}\approx
\frac{\Gamma(1/4)}{2\pi}\frac{g_1(\lambda^*) e^{\tau f_w(\lambda^*)}}{[2\tau f''_w(\lambda^*)]^{1/4}}\,\quad\text{as} ~\lambda^*\rightarrow
\lambda_\text{a}.
\label{P_S limit}
\end{equation}

 \section{Index of the notations}
 \label{notations}

 \begin{list}{\textbullet}{
 \setlength{\itemsep}{0pt} 
 \setlength{\parsep}{0pt}
 \setlength{\leftmargin}{3mm}
 \setlength{\topsep}{0pt}
 }
 \item $\gamma$ is the viscous drag.
 \item $k$ is the stiffness (spring constant) of the trap.
 \item $D=k_B T/\gamma$ is the diffusion constant.
 \item $\tau_\gamma=\gamma/k$ is the viscous relaxation time of the
 trap, which is introduced in \eref{Langevin-1}.
 \item $\tau_0$ is the correlation time of trap modulation, which is  
 introduced in \eref{Langevin-2}.
 \item $\tau_c$ is defined in \eref{mu}.
 \item $\theta$ and $\delta$ are  given by \eref{theta-delta}.
 \item $\alpha$ is defined by \eref{alpha}, and related to $\theta$ and
 $\delta$ by \eref{alpha-theta-delta}.
 \item $a$ is defined in \eref{nu-1}.
 \item $\mu(\lambda)$ is given by \eref{mu}.
 \item $\nu(\lambda)$ is given by \eref{nu-1} and \eref{nu-2}.
 \item $g(\lambda)$ is given by \eref{g}.
 \item $f_w(\lambda)$ is defined in \eref{f_w}.
 \item $b_\pm$ are defined by \eref{b_pm}.
 \item $\lambda_\pm$ are defined by \eref{lambda_pm}.
 \item $\lambda^*$ is given by \eref{lambda*}.
 \item $\lambda_i$ with $i\in\{\text{a},\text{b},\text{c},\text{d}\}$,
 are the positions of the branch points of $g(\lambda)$ and are given
  by Eqs.~\eqref{lambda-a}--\eqref{lambda-d}.
 \item $h_s(w)=f_w(\lambda^*)$ is given by \eref{hs(w)}.
 \item $h_i(w)=f_w(\lambda_i)$ with
 $i\in\{\text{a},\text{b},\text{c},\text{d}\}$, are given by
 Eqs.~\eqref{ha(w)}--\eqref{hd(w)}.
 \item $w^*_i$ with
 $i\in\{\text{a},\text{b},\text{c},\text{d}\}$, are given
 by \eref{wstar}.
 \item $\widetilde{g}(\lambda_i)$ with
 $i\in\{\text{a},\text{b},\text{c},\text{d}\}$, are defined by \eref{g-ns}.
 \item $\theta_{c_1}$, $\theta_{c_2}$, and $\theta_{c}$ are given by
 Eqs.~\eqref{theta_c1}, \eqref{theta_c2}, and \eqref{theta_c}
 respectively.
 \end{list}

\end{document}